\documentclass[a4paper,12pt]{article}
\pdfoutput=1
\usepackage{graphicx, rotating}
\usepackage{hyperref}
\usepackage{slashed}
\usepackage{ifpdf}

\ifx\pdfoutput\undefined
\usepackage[dvips,bookmarks=false]{hyperref}	% This is for arXiv.org
\else
\usepackage{hyperref}	% This is for pdftex
\fi
\hypersetup{colorlinks,bookmarksopen,bookmarksnumbered,citecolor=verdes,
linkcolor=blus,pdfstartview=FitH,urlcolor=rossos}
\def\hhref#1{\href{http://arxiv.org/abs/#1}{#1}} % in bibliography
\usepackage{amsfonts}
\usepackage{amsmath}
\usepackage{slashed}

\newcommand{\beq}{\begin{equation}}
\newcommand{\eeq}{\end{equation}}
\newcommand{\fig}[1]{~\ref{fig:#1}}

\newcount\Mac  \Mac=1  % devo mettere Mac=1 se sto lavorando sul file Mac
\newcommand{\ifMac}[2]{\ifnum\Mac=1 #1 \else #2 \fi}
\def\putps(#1,#2)(#3,#4)#5#6{\ifnum\Mac=1 \put(#1,#2){\special{picture #5}}
\else  \put(#3,#4){\includegraphics{#6}} \fi}

\newcommand{\One}{\hbox{1\kern-.24em I}}

\newcommand{\GeV}{\,{\rm GeV}}

\newcommand{\eq}[1]{~{\rm (\ref{eq:#1})}}

\newcommand {\leff}{\mathcal L_{\rm eff}}
\newcommand*{\isot}[2]{$\phantom{}^{#1}$#2}

\newcommand{\lascia}[1]{}
\makeatletter

\def\art{\@ifnextchar[{\eart}{\oart}}
\def\eart[#1]#2#3#4#5#6{{\rm #2}, {#3 #4} {\rm (#6) #5} [arXiv:{\hhref{#1}}]}

\def\hepart[#1]#2{{\rm #2, [arXiv:\hhref{#1}]}}
\newcommand{\oart}[5]{{\rm #1}, {#2 #3} {\rm (#5) #4}}

\numberwithin{equation}{section}

\newcounter{alphaequation}[equation]
\def\thealphaequation{\theequation\hbox to
0.6em{\hfil\alph{alphaequation}\hfil}}
\def\eqnsystem#1{
\def\@eqnnum{{\rm (\thealphaequation)}}
\def\@@eqncr{\let\@tempa\relax \ifcase\@eqcnt \def\@tempa{& & &} \or
  \def\@tempa{& &}\or \def\@tempa{&}\fi\@tempa
  \if@eqnsw\@eqnnum\refstepcounter{alphaequation}\fi
\global\@eqnswtrue\global\@eqcnt=0\cr}
\refstepcounter{equation} \let\@currentlabel\theequation \def\@tempb{#1}
\ifx\@tempb\empty\else\label{#1}\fi
\refstepcounter{alphaequation}
\let\@currentlabel\thealphaequation
\global\@eqnswtrue\global\@eqcnt=0 \tabskip\@centering\let\\=\@eqncr
$$\halign to \displaywidth\bgroup \@eqnsel\hskip\@centering
$\displaystyle\tabskip\z@{##}$&\global\@eqcnt\@ne
\hskip2\arraycolsep\hfil${##}$\hfil& \global\@eqcnt\tw@\hskip2\arraycolsep
$\displaystyle\tabskip\z@{##}$\hfil
\tabskip\@centering&\llap{##}\tabskip\z@\cr}

\def\endeqnsystem{\@@eqncr\egroup$$\global\@ignoretrue} \makeatother

\def\circa#1{\,\raise.3ex\hbox{$#1$\kern-.75em\lower1ex\hbox{$\sim$}}\,}

\usepackage{multicol}
\usepackage{color}
\definecolor{rosso}{cmyk}{0,1,1,0.4}
\definecolor{rossos}{cmyk}{0,1,1,0.55}
\definecolor{rossoc}{cmyk}{0,1,1,0.2}
\definecolor{blu}{cmyk}{1,1,0,0.3}
\definecolor{blus}{cmyk}{1,1,0,0.6}
\definecolor{bluc}{cmyk}{1,1,0,0.1}
\definecolor{verde}{cmyk}{0.92,0,0.59,0.25}
\definecolor{verdec}{cmyk}{0.92,0,0.59,0.15}
\definecolor{verdes}{cmyk}{0.92,0,0.59,0.4}
\definecolor{grigio}{cmyk}{0,0,0,0.07}
\definecolor{rosa}{cmyk}{0,0.1,0.1,0.02}
\definecolor{rosino}{cmyk}{0,0.05,0.05,0.02}
\definecolor{rosas}{cmyk}{0,0.3,0.25,0.05}
\definecolor{celeste}{cmyk}{0.1,0,0,0.02}
\definecolor{giallino}{cmyk}{0,0,0.4,0.02}
\definecolor{rosso}{cmyk}{0,1,1,0.4}
\definecolor{rossos}{cmyk}{0,1,1,0.55}
\definecolor{rossoc}{cmyk}{0,1,1,0.2}
\definecolor{blu}{cmyk}{1,1,0,0.3}
\definecolor{bluc}{cmyk}{1,1,0,0.1}
\definecolor{blucc}{cmyk}{0.7,0.5,0,0}
\definecolor{viola}{cmyk}{0,1,0,0.6}
\definecolor{viola2}{cmyk}{0,1,0.2,0.6}
\definecolor{verde}{cmyk}{0.92,0,0.59,0.25}
\definecolor{verdec}{cmyk}{0.92,0,0.59,0.15}
\definecolor{verdes}{cmyk}{0.92,0,0.59,0.4}
\definecolor{verdino}{cmyk}{0.12,0,0.09,0.05}
\definecolor{giallo}{cmyk}{0,0,1,0}
\definecolor{gialloverde}{cmyk}{0.44,0,0.74,0}

\newfam\rsfsfam
\def\mathscr#1{{\fam\rsfsfam\relax#1}}

\newcommand{\unit}[1]{\mathrm{\; #1}}
\newcommand{\cogent}{{\sc CoGeNT}}
\newcommand{\xenon}{{\sc Xenon100}}

\begin{document}
% IFUP-TH/2011-1\hfill CERN-PH-TH/2010-XX
\color{black}
\vspace{0.5cm}
\begin{center}
{\Huge\bf\color{rosso}Can CoGeNT and DAMA\\[3mm] Modulations Be Due to Dark Matter?}\\
\bigskip\color{black}\vspace{0.6cm}
{{\large\bf Marco Farina$^{a}$,    Duccio Pappadopulo$^{b}$,\\[2mm]
 Alessandro Strumia$^{c,d}$, Tomer Volansky$^{e,f}$}
} \\[7mm]
{\it  (a)  Scuola Normale Superiore and INFN, Piazza dei Cavalieri 7, 56126 Pisa, Italia}\\[3mm]
{\it  (b) Institut de Th\'eorie des Ph\'enom\`enes Physiques, EPFL,  CH--1015 Lausanne, Switzerland}\\[3mm]
{\it  (c) Dipartimento di Fisica dell'Universit{\`a} di Pisa and INFN, Italia}\\[3mm]
{\it  (d) National Institute of Chemical Physics and Biophysics, Ravala 10, Tallinn, Estonia}\\[3mm]
{\it  (e) Berkeley Center for Theoretical Physics, Department of Physics, \\University of California, Berkeley, CA 94720, USA}\\[3mm]
{\it  (f) Theoretical Physics Group, Lawrence Berkeley National Laboratory, \\
Berkeley, CA 94720, USA}\\[3mm]
%{\it CERN, PH-TH, CH-1211, Geneva 23, Switzerland}\\[3mm]
\end{center}
\bigskip
\centerline{\large\bf\color{blus} Abstract}
\begin{quote}
We explore the dark matter interpretation of the anomalies claimed by the DAMA and \cogent\ experiments, in conjunction with the various null direct-detection experiments.  An independent analysis of the \cogent\ data is employed and several experimental and astrophysical uncertainties are considered.  Various phenomenological models are studied, including isospin violating interactions, momentum-dependent form factors,  velocity-dependent form factors, inelastic scatterings (endothermic and exothermic) and channeling.
We find that the severe tension between the anomalies and the null results can be ameliorated but not eliminated, unless extreme assumptions are made.
%Nonetheless, the \cogent\ signal (but not DAMA) is found consistent with the null experiments under the simple assumption of  mild channeling. \color{black} \xxx{remove this statement?}
\end{quote}

\newpage

\tableofcontents

%%%%%%%%%%%%%%%%
\section{Introduction}
%%%%%%%%%%%%%%%%

Significant on-going efforts are being made to directly search for  Weakly Interacting Massive Particles (WIMPs).  An interesting and widely accepted signature of WIMPs is the annual modulation of their interaction rate, arising from the relative motion of the Earth around the Sun.
Strikingly,  both the DAMA~\cite{DAMA} and \cogent~\cite{cogent,Aalseth:2011wp} collaborations observe anomalous modulating events that may be interpreted as arising from interactions of spin-independent Dark Matter (DM).  As we discuss below, the two measurements are, to some extent,  consistent with each other, and point to a surprisingly low DM mass, of order a few GeV.

In contrast to the positive signals of DAMA and \cogent, several other experiments find no evidence for DM.  Most notably, the CDMS~\cite{CDMS}, {\sc Xenon}10~\cite{Angle:2009xb} and the \xenon~\cite{Xenon100} collaborations seem to disfavor  the parameter space indicated by DAMA and \cogent.
It is natural to ask, therefore, what possible systematic effects and/or DM properties can resolve the tension between the positive and null results?

One noticeable difference between DAMA/\cogent{} and the null experiments, is that the latter veto electronic interactions while attempting to collect only  nuclear recoil events.  It is conceivable that the anomalous signals arise from such electronic recoils, a possibility that would explain away the existing tension.   A model of this type was considered in~\cite{Kopp:2009et,Feldstein:2010su} prior to the recent \cogent\ measurement~\cite{Aalseth:2011wp} and it remains to be seen whether this possibility is theoretically feasible.
In this paper we pursue a different direction and study the viability of nuclear recoils of spin independent DM, as an explanation to the positive signals.  Several handles can, in principle, ameliorate the tension with the null results.  From the DM perspective one may consider,
\begin{itemize}
\item Inelastic scattering (endothermic or exothermic)~\cite{TuckerSmith:2001hy,Graham:2010ca,Cline:2010kv}.
\item Isospin-violating couplings~\cite{Kurylov:2003ra, Giuliani:2005my,Cotta:2009zu,Chang:2010yk,Kang:2010mh,Feng:2011vu,DelNobile:2011je}.
\item Velocity suppressed interactions.
\item Momentum dependent scattering~\cite{Feldstein:2009tr,Chang:2009yt}.
\item Resonant scattering~\cite{Bai:2009cd}.
\end{itemize}
In addition there are uncertainties that may significantly change the expected scattering rates in various experiments:
\begin{itemize}
\item Astrophysical uncertainties~\cite{Smith:2006ym,Bruch:2008rx,Fairbairn:2008gz,Kuhlen:2009vh,Fox:2010bz,Fox:2010bu,Savage:2006qr,Lisanti:2010qx,Arina:2011si}.
\item Possible  channeling effects~\cite{Drobyshevski:2007zj,Bernabei:2007hw} (see however~\cite{channeling}).
\item Uncertainties in the Xenon scintillation function at low recoil energies or in the statistical treatment of the background events (for a discussion, see e.g.~\cite{Collar:2011wq}).
\end{itemize}
A fully systematic analysis that takes into account all of the theoretical and experimental uncertainties above is hard to attain and we do not attempt here.   Instead,  we separately study the effects  of  most  of the above possibilities on the DM interpretation of the DAMA and \cogent\ results, as well as the null experiments.

Our goal is two-fold.  Primarily, we aim at understanding to what extent the DAMA and \cogent\ results are consistent with each other, and with the other null experiments (for related works, see~\cite{Hooper:2011hd}).   In addition, we study what is required from the theoretical point of view and what needs to be assumed on the experimental side, in order to ameliorate the tension.  Along the way, we reanalyze the \cogent\ modulation data.

The paper is organized as follows.  In Section~\ref{th} we summarize the standard computation of the expected signals and shortly discuss the DM velocity distributions considered here.
In Section~\ref{duccio} we describe the experimental data we use, dwelling on the various uncertainties and discussing their influence on the fits.
Our results are presented in Section~\ref{res}.  Here we follow the analyses endorsed by the various experimental collaborations
and  explore whether velocity distributions or any of the DM scenarios mentioned above can explain the signals compatibly with the bounds.
We conclude in section~\ref{concl}.

%%%%%%%%%%%%%%%%%%%%%
\section{Formalism}\label{th}
%%%%%%%%%%%%%%%%%%%%%

The direct detection rate for DM--nucleus scattering at a given experiment is given by
\begin{equation}\label{rate}
\frac{d R}{d E_{\rm R}}=N_T\frac{\rho_\odot}{M_{\rm DM}}\int_{|\vec v|>v_{\min}} d^3v\,v\,f_\oplus(v,t)\,\frac{d\sigma}{d E_{\rm R}}\,,
\end{equation}
where $N_T$ is the number of target nuclei per unit mass of the detector, $M_{\rm DM}$ is the DM mass,  $\rho_\odot$ is the local DM density (that we assume to be equal to $0.3\unit{GeV/cm^3}$)
and $f_\oplus(v,t)$ is the DM velocity distribution in the Earth frame, to be discussed below.
 We denote $\sigma$ to be the DM--nucleus scattering cross section, which we take to be of the form,
\begin{equation}
\label{eq:dsigmadE}
\frac{d\sigma}{dE_{\rm R}} = \frac{m_N\sigma_n}{2 v^2\mu_n^2} \frac{[f_p Z + f_n (A-Z)]^2}{f_n^2} F_N^2(q)F_{\rm DM}^2(q,v)\,.
\end{equation}
Here $m_N$ is the nucleus mass, $\mu_n$ is the DM-nucleon reduced mass, $f_p$ ($f_n$) is the coupling strength to the proton (neutron) and $F_N(q)$ is the nucleus form factor. Throughout this work we use the Helm form factor given in~\cite{Lewin:1995rx}.   The DM form factor, $F_{\rm DM}(q,v)$, is a velocity and/or momentum dependent contribution to the cross section which may exist, depending on the DM coupling.  Below we consider several possibilities for its form.
%(which may depend on $E_R$ and $v$)
Finally, $v_{\min}$ appearing in Eq.~(\ref{rate}) is the minimal DM velocity needed for a scattering with a recoil energy $E_{\rm R}$ to occur. In the elastic scattering case, it is given by
\begin{equation}
v_{\min}=\sqrt{\frac{m_N E_{\rm R}}{2\mu^2}}\,.
\end{equation}
 In Section~\ref{sec:iDM} we consider inelastic scattering, in which case the required minimal velocity  depends further on the mass splitting, $\delta = M'_{\rm DM} - M_{\rm DM}$, of the recoiling particles:
\begin{equation}
\label{vminidm}
v_{\min}=\frac{1}{\sqrt{2m_N E_{R}}}\left| \frac{m_NE_R}{\mu}+\delta\right|\,.
\end{equation}

%\bigskip

The scattering rate is expected to be time-dependent, exhibiting a maximum at the beginning of June (when the Earth moves against the DM wind) and a minimum at the beginning of December (when the Earth moves along the DM wind). It is thus useful to define a modulated rate as
\begin{equation}
\frac{d R_{\rm mod}}{d E_{\rm R}}=\frac{1}{2}\left[\frac{d R}{d E_{\rm R}}({\rm{2 ~June}})-\frac{d R}{d E_{\rm R}}({\rm{2 ~December}})\right].
\end{equation}
Both the modulated and unmodulated rates can be confronted with the results of an experiment only after accounting for all expected backgrounds and experimental efficiencies. The modulated rate is, in particular, independent of any time-independent background which can contaminate the DM signal.

\begin{figure}[t]
$$\includegraphics[width=0.45\textwidth]{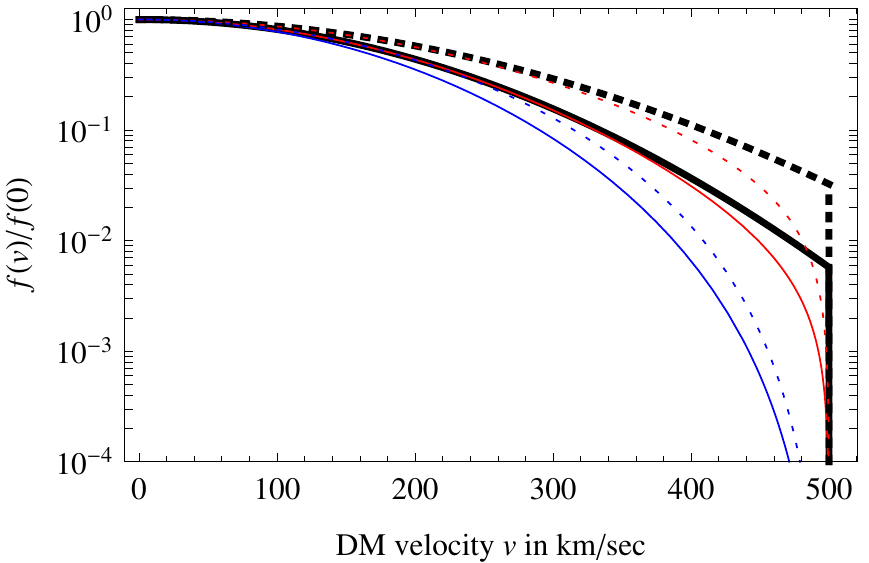}\qquad
\includegraphics[width=0.47\textwidth]{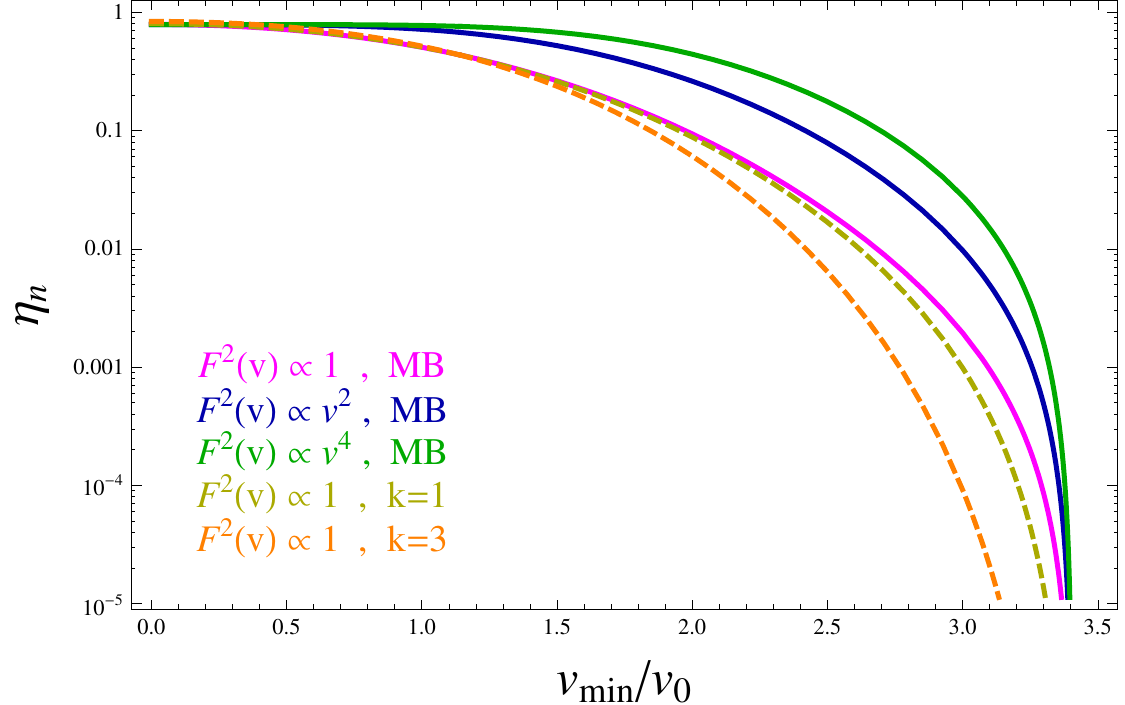}$$
\caption{\em {\bf Left:} Dark Matter velocity distributions:  the Maxwell-Boltzmann with sharp cutoff at $v=v_{\rm esc} = 500\,{\rm km/s}$ (thick black curve), the distribution, Eq.~\eqref{eq:fk}, with a smooth cutoff computed for $k=1$ (red) and $k=3$ (blue).
Two different values of $v_0$ are shown: $220~{\rm km/s}$ (solid) and $270~{\rm km/s}$ (dotted).
{\bf Right:} The function $\eta_n$ defined in Eq.~\eqref{eq:eta} as a function of the minimal DM velocity, $v_{\rm min}$, normalized to the mean DM velocity, $v_0$.   As discussed in the text, this function captures the effect of velocity-dependent form factor on the total scattering rate.  We plot $F^2_{\rm DM}(v)=1, v^2/v_0^2, v^4/v_0^4$ (magenta, blue and green respectively) for the Maxwell-Boltzmann distribution.  For completeness we also show the corresponding function for the $\tilde f_k$ velocity distribution defined in Eq.~\eqref{eq:fk} with a trivial form factor.  $k=1$ (dashed yellow) and $k=3$ (dashed orange) are shown.   For all lines, we use, $v_0=220$ km/sec, $v_{\rm esc}=500$ km/sec, and take the Earth velocity on June 2nd.
\label{fig:veldistr} \label{fig:eta}}
\end{figure}

%%%%%%%%%%%%%%
\subsection{Dark Matter velocity distributions}\label{DMv}
%%%%%%%%%%%%%%
The DM   interaction rate with  nuclei depend on the velocity distribution $f_\oplus$ of DM as in eq.~(\ref{rate}).  Since this distribution has not been measured, it adds its own source of uncertainty to the DM interpretation of direct detection experiments.   This topic has been extensively studied in the literature, where several possibilities such as streams or velocity substructure were considered~\cite{Bruch:2008rx,Fox:2010bu,Savage:2006qr,Lisanti:2010qx}.   While a systematic study of astrophysics uncertainties and its prospects for ameliorating the  experimental tension is beyond the scope of this paper,  we consider  few  motivated distributions to demonstrate the possible variation in the fits.  We stress that it is conceivable for significantly different conclusions to be drawn with the use of less conventional or more exotic possibilities, as was demonstrated prior to the new \cogent\ result in, e.g.~\cite{Fox:2010bu}.

The velocity distribution in our local frame, $f_\oplus(v,t)$ defined in Eq.~\eqref{rate}, is conveniently expressed in terms of the velocity distribution in the galactic frame, $\tilde f(v)$, through
\begin{equation}
\label{eq:fearth}
f_\oplus(v,t) = \tilde f(v+v_\oplus(t); v_0, v_{\rm esc})\,.
\end{equation}
Here $v_\oplus$ is the relative motion of the Earth with respect to the galactic frame (see e.g.~\cite{Savage:2006qr} for further details).    $v_0$ is the root mean square velocity typically taken to be in the range, $220<v_0<270$ km/s, while $v_{\rm esc}$ is the escape velocity, in the range, $450< v_{\rm esc}< 650$ km/s~\cite{Smith:2006ym}.   Below, we  consider $v_0 = (220 ,270)\unit{km/s}$ and $v_{\rm esc} = (500, 600)\unit{km/s}$.

The DM  velocity distribution in the galactic frame is often assumed to be a Maxwell-Boltzmann (MB) sharply  cut off by a finite escape velocity,
\begin{equation}
\label{MB}
\tilde f_{\rm MB}( v; v_0,v_{\rm esc}) = \frac{1}{N_E} e^{-v^2/v_0^2}\Theta(v_{\rm esc}-v) \,,
\end{equation}
with $N_E = ({\rm erf}(z)-2z\exp(-z^2)\pi^{-1/2}) \pi^{3/2} v_0^3$ and $z=v_{\rm esc}/v_0$.      Above, we have explicitly denoted the dependence of the velocity distribution on $v_0$ and $v_{\rm esc}$.
The above MB distribution does not seem  to capture the results of $N$-body simulations~\cite{Fairbairn:2008gz,Kuhlen:2009vh,Vogelsberger:2008qb,Ling:2009eh}.  An improved ansatz for an isotropic velocity distribution is given by
\begin{equation}
\label{eq:fk}
\tilde f_k(v; v_0, v_{\rm esc})\propto \left[\exp\left(\frac{v_{\rm esc}^2-v^2}{kv_0^2}\right)-1\right]^k\Theta(v_{\rm esc}-v)\,
\end{equation}
for $1.5<k<3.5$~\cite{Lisanti:2010qx}.  The MB distribution is reobtained in the limit $k\rightarrow 0$.
These velocity distributions are plotted in Fig.\fig{veldistr}a.

The above distributions enter the scattering rate, Eq.~\eqref{rate}, through the function,\footnote{We use the notations of~\cite{Savage:2006qr}, however our function $\eta_n$ is chosen to be dimensionless, as opposed to the corresponding function defined in~\cite{Savage:2006qr}.}
 \begin{equation}
 \label{eq:eta}
 \eta_n\left(\frac{v_{\rm min}}{v_0},\frac{v_\oplus}{v_0},\frac{v_{\rm esc}}{v_0}\right) = \int_{|\vec v|>v_{\rm min}} d^3v~ f_\oplus(v,t) \left(\frac{v}{v_0}\right)^{-1}F_{\rm DM}^2(q,v)\,.
 \end{equation}
Note that the dependence on $v_\oplus$ and $v_{\rm esc}$ is implicit through the definition of $f_\oplus$ [see Eq.~\eqref{eq:fearth}].   In  Fig.~\ref{fig:veldistr}b we plot $\eta_n$ for  $F^2_{\rm DM}(q,v)=1, v^2/v_0^2, v^4/v_0^4$ with the Maxwell-Boltzmann distribution, as well as the trivial form factor for the $\tilde f_k$ velocity distribution defined in Eq.~\eqref{eq:fk}.   As can be seen, higher powers of velocity in the form factor imply a larger overall normalization and a larger sensitivity to higher minimal velocity with a sharper falloff.  We return to these features in $\eta_n$ in Section~\ref{sec:velocity}, where we study velocity-dependent form factors.

\section{Experiments}\label{duccio}

In this section we briefly summarize the data used to derive the allowed region for spin-independent DM scattering.  For each experiment we stress the various sources of uncertainties and our approach for dealing with them.  In most cases, we consider several possibilities, allowing for a conservative view of the tension between the positive and null experimental results.

\begin{table}[t]
\begin{center}
\begin{tabular}{ccc}
bin (keVee) & rate (${\rm cpd^{-1}\,kg^{-1}\,keVee^{-1}}$) & $\sigma_{\rm rate}$\\ \hline\hline
2--2.5&0.016&0.004\\
2.5--3&0.026&0.005\\
3--3.5&0.022&0.005\\
3.5--4&0.008&0.005\\
4--4.5&0.011&0.004\\
4.5--5&0.005&0.004\\
5--5.5&0.009&0.003\\
5.5--6&0.004&0.003\\
6--14&0.000&0.000\\
\hline\hline
\end{tabular}
\end{center}
\caption{\small\label{damadata}\em The data used in the fit to the DAMA modulated amplitude.}
\end{table}

\subsection{DAMA}
The DAMA experiment employs a NaI(Tl)  target and observes an 8.9$\sigma$ evidence for an annual modulation in its energy spectrum \cite{DAMA}. The modulation is present in the 2-6 keVee energy range and the time dependence of the rate is consistent with the hypothesis of DM scattering on nuclei. The DAMA experiment is able to measure only the fraction of
energy  that recoil nuclei deposit as scintillation. This fraction of the total recoil energy, known as the quenching factor, is taken to be $q_{\rm I}=0.09$ for Iodine.

Different groups report varying values for the quenching factor on Sodium.   The DAMA collaboration reports $q_{\rm Na}=0.30\pm 0.01$ averaging over recoil energies ranging from 6.5 to 97 keV.
 Ref.~\cite{Tovey:1998ex} finds $0.33\pm0.15$ between 4 to  11 keV while ref.~\cite{chagani} finds $0.252\pm0.065$ around 10 keV.
The relevance of $q_{\rm Na}$ resides in the fact that larger values ameliorate the apparent tension between DAMA and the null experiments.  Indeed larger $q_{\rm Na}$
 implies lower recoil energies at DAMA and consequently favors smaller WIMP masses.
Following many previous works, we adopt a conservative estimation of the uncertainties, assuming $q_{\rm Na}=0.3\pm 0.1$.

An additional source of uncertainty arises due to the crystalline nature of the target material.  It is possible that some (experimentally unknown) fraction of the ion scatterings occur parallel to a symmetry axis (channeled events),  depositing the entire energy in scintillation.  For such events the quenching factor is effectively 1 and hence a sizable fraction of  channelled events may  significantly alter the direct detection predictions~\cite{Drobyshevski:2007zj}.     Recent theoretical results,  suggest that the channeling fraction in NaI is negligible~\cite{channeling}.
In order to be {\em over}-conservative, in Section~\ref{sec:channeling} we also perform fits in which we allow an energy independent channeling fraction.

The spin independent fit to the DAMA modulated rate allows two qualitatively different best-fit regions: one around $M_{\rm DM}\approx 80~{\rm GeV}$ with $\sigma\approx 10^{-41}~{\rm cm^2}$ due to scattering on iodine ($A=127$, $Z=53$) and one around $M_{\rm DM}\approx 10~{\rm GeV}$ with $\sigma\approx 10^{-40}~{\rm cm^2}$ due to scattering on sodium ($A=23$, $Z=11$). The former region is firmly excluded by many other experiments, most notably XENON10/100. The latter region, while still disfavored by other null searches, is not as badly excluded due to many experimental uncertainties and due to the general difficulty of direct detection searches to deal with low recoil energy scatterings. In the following we will focus on this low mass region, being also the only possibility to reconcile DAMA with the results of {\cogent}.

To fit the modulated signal at DAMA we build a simple $\chi^2$ using the content of the first 8 bins in Fig. 9 of \cite{DAMA}. We use a single bin for the modulated rate in the energy range of 6 to 14 keVee. The data we use are shown in Table~\ref{damadata}.
DAMA collaboration also reports (see fig.~1 of \cite{DAMAold}) a measurement of its non-modulated rate. This can be used to define an upper bound on the signal, following for instance the procedure of \cite{Papucci:2009gd}. Unless stated otherwise, we will not use this bound in the fits.

To compare the data with the theoretical hypothesis we introduce a finite detector resolution, parameterizing the energy smearing through a gaussian with energy dependent width given by,
\begin{equation}
\sigma_{\rm DAMA}\left(\frac{E}{\rm keVee}\right)=0.448\sqrt{\frac{E}{\rm keVee}}+9.1\times 10^{-3}\frac{E}{\rm keVee}\,.
\end{equation}

 \subsection{CoGeNT}

\begin{table}[t]
\begin{center}
\begin{tabular}{cccccc}
 & $N$ &$\delta N/N$& $E_{\rm peak}$ (keV)&$\sigma_{E_{\rm peak}}$&$T_{1/2}$ (days) \\ \hline\hline
\isot{73}{As} & 12.7& 0.33 & 1.414 & 0.078 & 80  \\
\isot{68}{Ge} & 639 &  0.01&1.298 & 0.077 & 271  \\
\isot{68}{As} & 52.8& 0.05 & 1.194 & 0.076 & 271  \\
\isot{65}{Zn}& 211 & 0.02 &1.096 & 0.076 & 244  \\
\isot{56}{Ni}& 1.53& 0.23 & 0.926 & 0.075 & 6  \\
\isot{56,58}{Co}& 9.44& 0.45 & 0.846 & 0.074 & 71  \\
\isot{57}{Co} & 2.59 & 3.81& 0.846 & 0.074 & 271  \\
\isot{55}{Fe}& 44.9 & 0.12 & 0.769 & 0.074 & 996 \\
\isot{54}{Mn} & 21.1& 0.09 & 0.695 & 0.074 & 312  \\
\isot{51}{Cr}& 2.94& 0.15 & 0.628 & 0.073 & 28  \\
\isot{49}{V}& 14.9& 0.12 & 0.564 & 0.073 & 330 \\
\hline\hline
\end{tabular}
\end{center}
\caption{\label{Lshellcogent}\em The data  used to extract the L-shell backgrounds. $N$ is the total number of decays expected from a given isotope from the beginning of the {\cogent} data taking to the end of time, with $\delta N/N$ being its relative error.  $E_{\rm peak}$ is the central value for the corresponding binding energy while $\sigma_{E_{\rm peak}}$ is the energy resolution.  $T_{1/2}$ is the half-life of the relevant isotope.}
\end{table}

The {\cogent} experiment uses a Germanium ($A\approx76$, $Z=32$) detector and takes data in the Soudan Underground Laboratory (SUL). We employ the latest data release~\cite{Aalseth:2011wp}, which are the result of 442 live days of data-taking from January 4 2010 to March 6 2011 on a 0.33 kg Germanium target. The new  data confirm the presence of an exponential distribution of events between 0.5 and 1.5 keV which is not accounted for by any known background. A time analysis of the same data also shows evidence of an annual modulation in the 0.5--3.0 keV range, which could be interpreted as  evidence for DM interacting with the detector.

To extract the rate and the modulation to be used in the fit, we use the time-stamped raw data obtained from the {\sc CoGeNT} collaboration \cite{datacollar}.
 The relevant region between 0.4 and 3.5 keV is contaminated by events due to the electron capture (EC) decay of cosmogenically activated elements in the detector. In a given time window these events are expected to show up as peaks centered around  the L-shell binding energy of the daughter of the decaying nucleus. The width of the peaks is related to the detector resolution and it's amplitude to the number of active isotopes in the detector. The amplitude of the peak is thus expected to decay in time with the half--life of the relative isotope. In Table~\ref{Lshellcogent} we show the parameters needed to describe these background events. It is clear that due to the time dependence of such backgrounds, their subtraction is of primary importance before attempting any time-analysis of the signal.

\begin{figure}[t]
$$\includegraphics[width=0.45\textwidth]{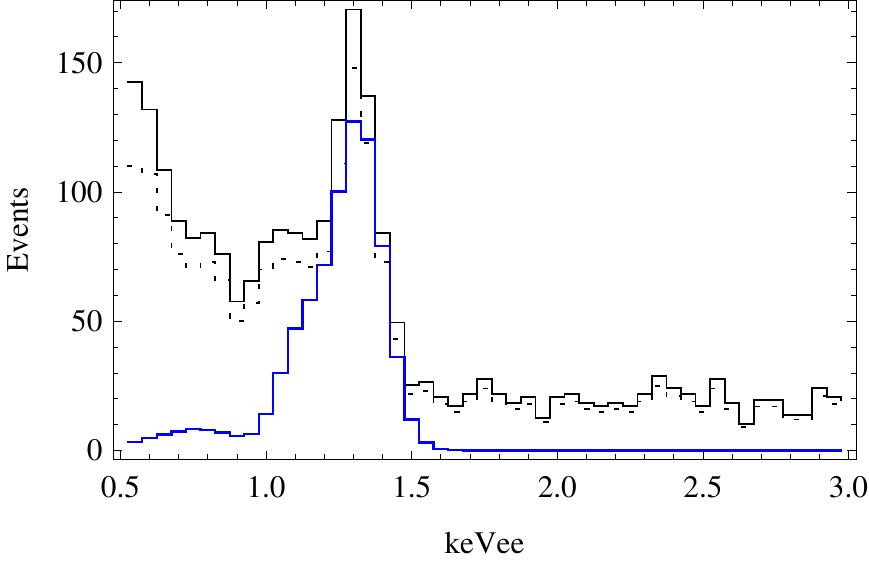}\quad\includegraphics[width=0.45\textwidth]{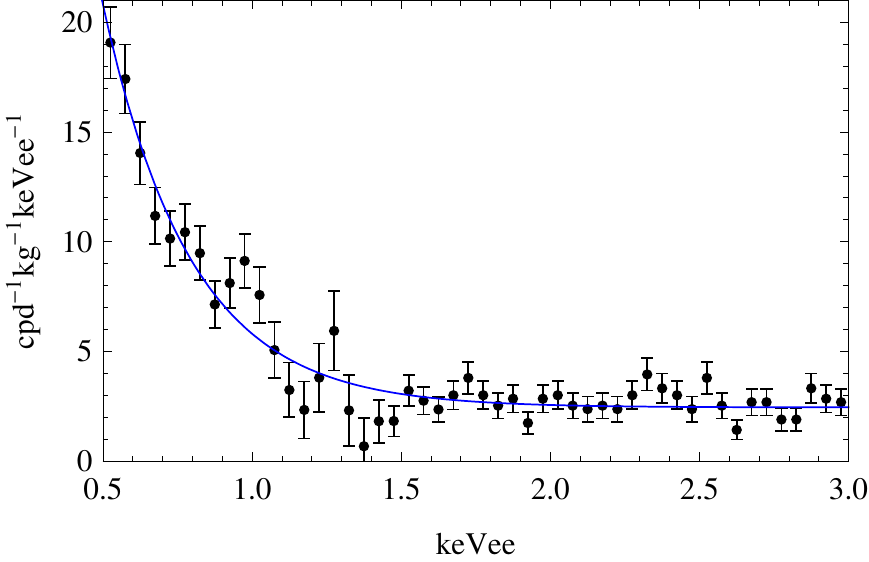}$$
\caption{\label{cogentunmod} \em {\bf Left:} distribution of the events observed during 442 days of live-data taking (black dashed). The black line is the result of efficiency unfolding. The blue contour shows the contribution from L-shell EC lines calculated from the data in tab.~\ref{Lshellcogent}. {\bf Right:} {\cogent}{} non-modulated rate after the subtraction of L-shell lines. The blue line is the result of an exponential+constant fit to the rate. In both plots we use a uniform (0.1 keVee) binning.}
\end{figure}

In Fig.~\ref{cogentunmod} we show the spectrum of events observed at {\cogent} as well as the rate after efficiency unfolding. As shown, the residual events after subtracting the L-shell peaks, are well fitted by an exponential plus constant function $A e^{-B E}+C$, with $E$ in keVee. We find,
\begin{equation}\label{bestfitcog}
A=100.4,\quad B=3.4,\quad C=2.4,\quad \chi^2/{\rm d.o.f.}=57/(50-3)\,.
\end{equation}
\begin{figure}[t]
$$\includegraphics[width=0.5\textwidth]{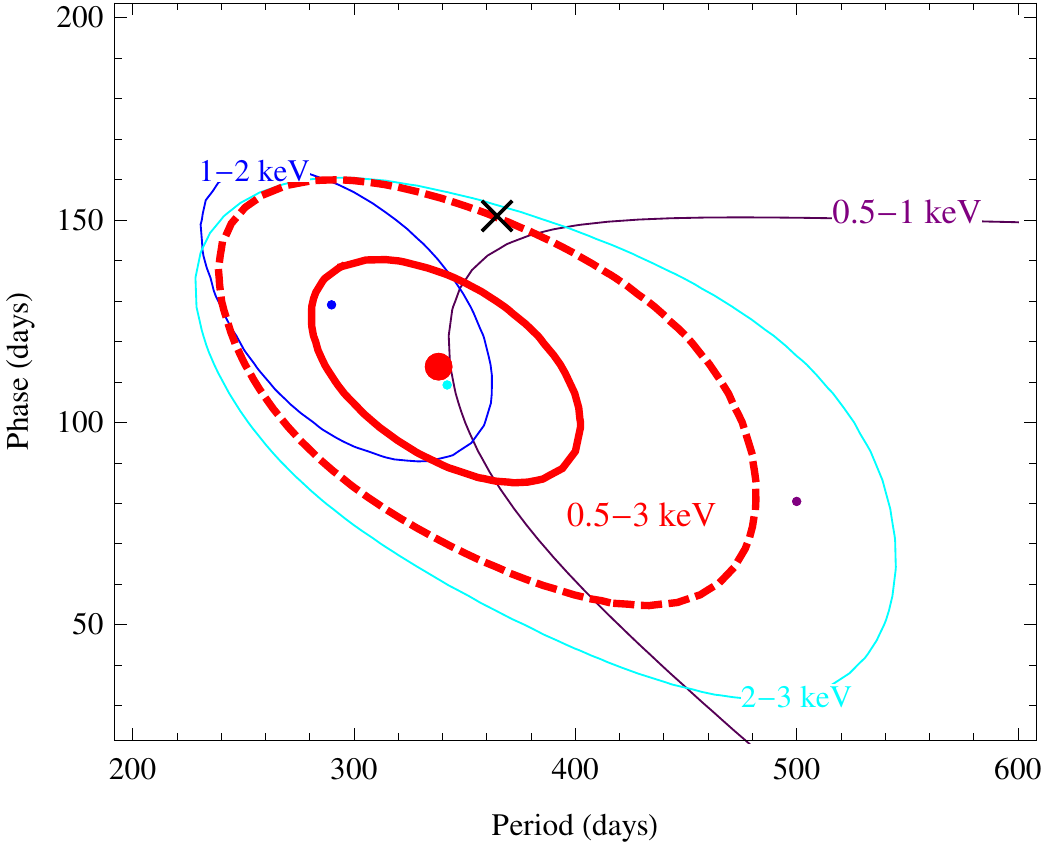}$$
\caption{\em\label{modcogTphi} Best fit for the period $T$ and phase $\phi$ of the  \cogent{}signal in the low energy bins.
The continuous lines are the 1$\sigma$ contours.  For the large 0.5-3 keVee bin, we also show the 2$\sigma$ one (dashed). The black cross indicates the prediction of  the DM hypothesis.
}
\end{figure}
For the time-analysis we use data up to 6 keVee binned into 5 energy bins: 0.5-1, 1-2, 2-3, 3-4.5, 4.5-6 keVee. Each bin is further divided into 15 time bins of 30 days each. For every energy--time bin we subtract the estimated number of events from L-shell EC and apply a correction to account for the offline time. Since in the first bin the efficiency is not constant, we choose not to unfold it to avoid loosing more statistics. A first test to the DM hypothesis is to check whether the signal modulates with a yearly period peaking on $2^{\rm nd}$ of June. We thus fit the signal to the function
\begin{equation}\label{cogentfit}
\mathcal B+\mathcal S \cos\frac{2\pi (t-\phi)}{T}\,,
\end{equation}
and check the consistency with the hypothesis $T=365$ days and $\phi=152$ days with the use of a $\chi^2$ analysis, marginalizing over $\mathcal B$ and $\mathcal S$. The results of the fit are shown in Fig.~\ref{modcogTphi}. The data show consistency with the DM hypothesis at the 2$\sigma$ level.
The signal deviates from the null (no-modulation) hypothesis, at the 2$\sigma$ level at most for each bin separately, while it exceeds this value (however not reaching 3-sigma) for the combined 0.5-3 keVee bin.    The deviation from the null hypothesis is demonstrated further in Fig.~\ref{spectrummod}, where the modulation spectrum is shown.

\begin{figure}[t]
$$\includegraphics[width=0.3\textwidth]{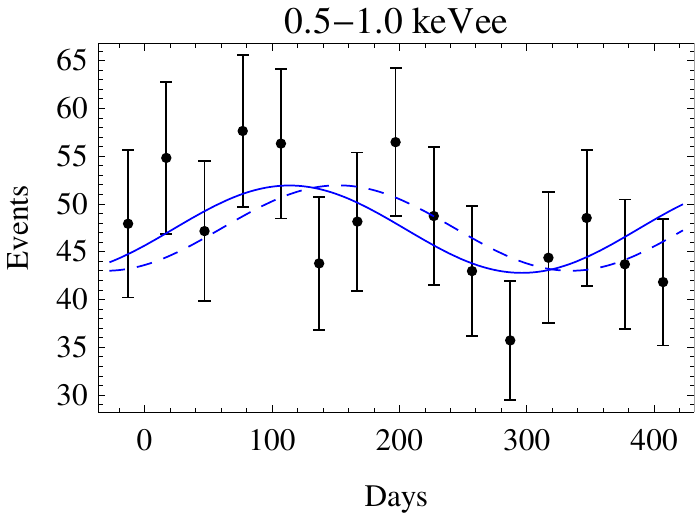}~~\includegraphics[width=0.3\textwidth]{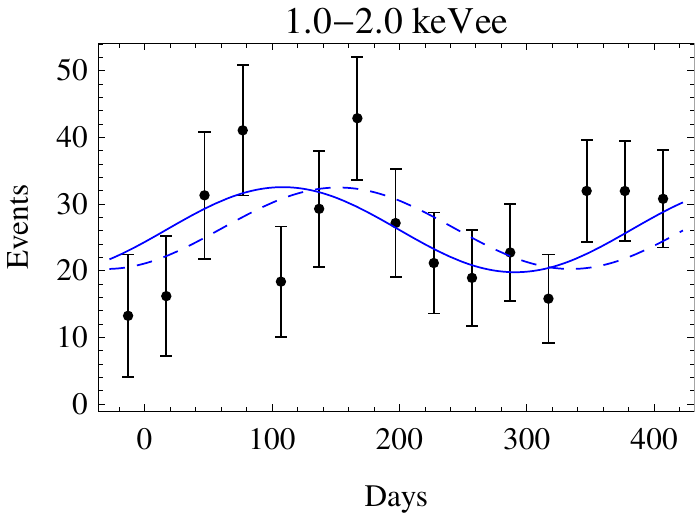}~~\includegraphics[width=0.3\textwidth]{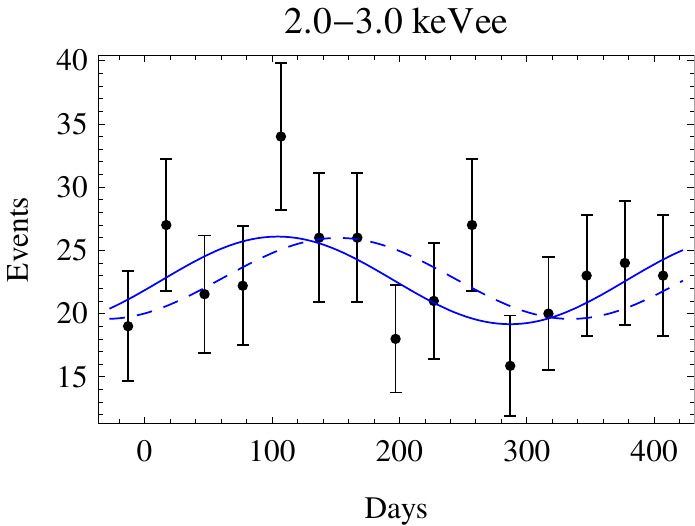}$$
\caption{\em\label{figmodanalysis} Time-analysis of the data in the bins of interest. In all the fits we keep $T=365$~days fixed. The blue full lines are obtained floating $\mathcal B$, $\mathcal S$ and $\mathcal \phi$. The blue dashed lines are the results of the fit fixing also $\phi=152$ days.}
\end{figure}

\begin{figure}[t]
$$\includegraphics[width=0.55\textwidth]{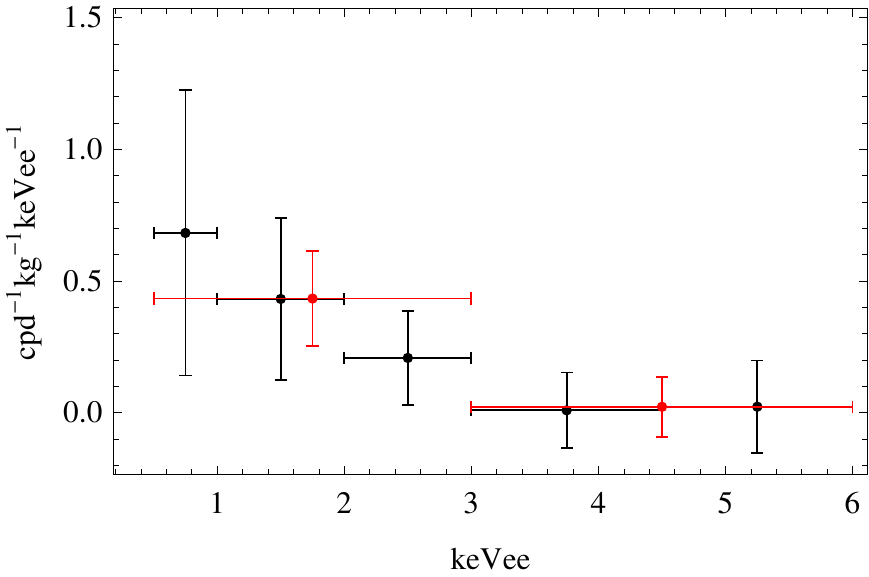}$$
\caption{\em\label{spectrummod} Spectrum of the modulation amplitude in the \cogent{} data assuming $T=365$ days and $\phi=152$ days. In red we show the result of fitting just two bins: 0.5-3 and 3-6 keVee. No efficiency correction is applied.}
\end{figure}

We thus fix $T$ to the value predicted by the DM hypothesis, 365 days, and  proceed to determine the best fit to the other parameters $\mathcal B$, $\mathcal S$ and $\phi$. The  results are summarized in Table~\ref{tabmodanalysis} and shown in Fig.~\ref{figmodanalysis}. The dashed curves in Fig.~\ref{figmodanalysis} are obtained by fixing both $T=365$ days and $\phi=152$ days. To extract the amplitude of the modulation $\mathcal S$ to be used in what follows, we marginalize over $\mathcal B$ to obtain the spectrum shown in Fig.~\ref{spectrummod}. In the two bins $0.5-3$ and $3-6$ keVee we get (in ${\rm cpd^{-1}kg^{-1}keVee^{-1}}$)
\begin{equation}
0.5-3\,{\rm keVee}:\,0.43\pm0.18,\quad\qquad 3-6\,{\rm keVee}:\,0.02\pm0.11
\end{equation}

Finally, we need to specify the quenching factor of Germanium in order to translate keVee energy into keV nuclear recoil energies. We use a Lindhard $k=2$ parametrization
\begin{equation}\label{eq:qGe}
\frac{E}{\rm keVee}=0.2 \left(\frac{E_{{R}}}{\rm keV}\right)^{1.12}.
\end{equation}

\begin{table}[t]
\begin{center}
\begin{tabular}{ccccc}
 $E$ (keVee)& $\mathcal B$ &$\mathcal S$& $\phi$& $\chi^2$\\ \hline\hline
$0.5-1.0$ & 47.3(47.6) & 4.6(3.4) & 114(152) & 7.9(9.1)\\
$1.0-2.0$ & 26.1(26.3) & 6.4(4.3) & 108(152) & 11.5(13.7)\\
$2.0-3.0$ & 22.6(22.8) & 3.5 (2.1)& 104(152) & 8.0(10.1)\\
$3.0-4.5$ & 34.2(34.2) &0.2(0.1)  &211(152) &15.7(15.7)\\
$4.5-6.0$ &46.9(47.0) & 1.9(0.3) &77(152)& 21.2(21.7)\\
\hline
$0.5-3.0$ & 97.2(97.7) & 15.0(10.7) & 112(152) &6.3(11.0)\\
$3.0-6.0$ & 82.5(82.7) & 2.7(0.7)& 81(152) &14.3(14.9)\\
\hline\hline
\end{tabular}
\end{center}
\caption{\label{tabmodanalysis}\em Summary of the time-analysis of the \cogent{} data. Everywhere $T=365$ is fixed. The numbers in parenthesis are obtained fixing $\phi=152$ days. The number of d.o.f. are 12 (13), in each energy bin.}
\end{table}

 \subsection{SIMPLE}
The SIMPLE experiment is a superheated liquid ${\rm C_2Cl\,F_5}$ droplet detector. Working in a manner similar to bubble chambers it looks for bubble nucleations induced by WIMPs. In particular, due to the presence of light ions such as fluorine ($A=19$, $Z=9$), it is sensitive to light mass WIMPs even with a relatively high threshold of 8 keV. Although it is not possible for this kind of experimental setup to measure events' energies, the threshold can be set precisely as the bubble nucleation depends on the temperature and pressure of the superheated liquid. Neutron induced recoils has confirmed the minimum threshold energy at 8 keV with a precision of 0.1 keV.

The SIMPLE collaboration has recently published new results \cite{Felizardo:2011uw} including new Stage 2 data which improves and merges the analysis of older (Stage 1) data.  As the merging can potentially introduce additional systematic uncertainties, we use only the Stage 2 data, with zero unidentified events and an exposure of 6.71 kg\,days. We set an exclusion limit using a Poissonian likelihood
\begin{equation}
\mathcal L_{\rm SIMPLE}= e^{-N_{\rm DM}}.
\end{equation}
Such a likelihood is considered to be a conservative choice that produces bounds compatible with the official one.

\subsection{XENON100}
{\xenon} is a two-phase Xenon ($A\approx131$, $Z=54$)  experiment which published results obtained from approximately $100$ live days of data acquisition in a fiducial volume of 48 kg. The consistency of the outcome with the background hypothesis allows to place strong constraint on the interaction of a WIMP with Xenon nuclei.

\begin{figure}[t]
$$\includegraphics[width=0.5\textwidth]{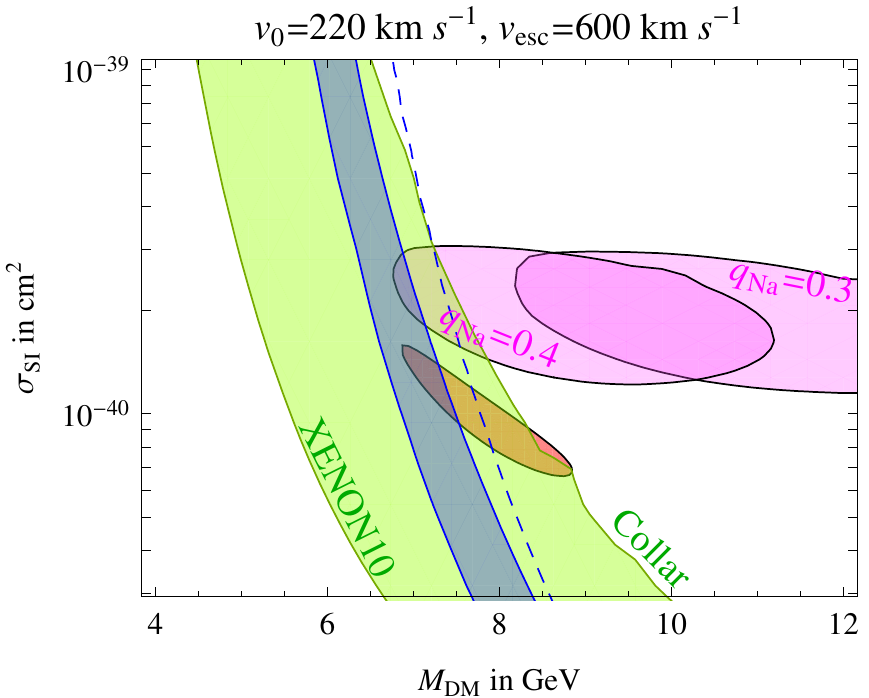}$$%~~\includegraphics[width=0.4\textwidth]{figs/pXe2.pdf}$$
\caption{\em\label{xenonbound}{\sc Xenon100/10} bound compared with the 3$\sigma$ favored region for DAMA (magenta) and \cogent\,(red). The blue band is the \xenon\, exclusion obtained by varying $\leff$ in its 1 $\sigma$ range. The blue dashed line shows the result of the conservative choice $\leff=0$ below $4\,{\rm keV}$. The light-green band represents how the 3$\sigma$ {\sc Xenon10} exclusion changes with respect to $\mathcal Q_y$ variations: on the left we use the choice of the {\sc Xenon10} collaboration and on the right we use the conservative estimate adopted in \cite{Collar:2011wq}. Two choices for the sodium quenching factor $q_{Na}$ are adopted for the DAMA fits.}
\end{figure}

As opposed to DAMA and {\cogent}, {\xenon} has a signal-to-background discrimination ability. This is achieved by comparing the primary scintillation signal (S1) to the ionization yield (S2), the relative magnitude of the latter being bigger for electronic recoils. The capability of {\xenon} to detect low mass WIMP scatterings crucially depends on the response function $\leff$ which, through the relation
\begin{equation}\label{scintillation}
S1(E_{{R}})=3.6~{\rm PE}\times E_{{R}}\times\leff,
\end{equation}
gives the number of photoelectrons (PE) in the S1 signal as a function of the recoil energy. Measurements of $\leff$ extend down to 3 keV \cite{Plante:2011hw} while extrapolation have to be used for lower values.  It was argued in~\cite{Collar:2011wq}, that the uncertainty on $\leff$ significantly influences the ability of \xenon\, to constrain light DM.  Below, we use the $\leff$ contours adopted by the {\xenon} collaboration and shown in Fig.~1 of~\cite{Xenon100}. % As a conservative measure we  also display the limits assuming $\leff=0$ below 4 keV.
To have a good signal-to-background discrimination, the lower {\xenon} threshold is fixed at 4 photoelectrons which corresponds to roughly 8~keV (depending on the precise choice of $\leff$). The relevance of the extrapolation of $\leff$ to lower energies has to do with the statistical nature of the scintillation process. Assuming the photoelectron generation process to be poissonian in nature, with a mean dictated by Eq.~(\ref{scintillation}), recoil occurring below threshold will have a non vanishing probability to generate an S1 signal above threshold. This tail is crucial to the constraining power of {\xenon} for low mass WIMPs\footnote{The {\xenon} analysis considers an S2 threshold of 300 photoelectrons. Values of $\langle$S1$\rangle$ below threshold may generate, through a statistical fluctuation,  an S1 signal above threshold but may fail to pass the S2 cut. Following \cite{gondoloxenon} we ignore recoils giving $\langle$S1$\rangle\leq1$ PE.}.  We have checked and found that assuming a different statistical behaviour of the S1-generation process (binomial for instance, see \cite{Collar:2011wq}) does not alter our conclusions.   We show the \xenon\, bound and its uncertainty in Fig.~\ref{xenonbound}. The blue band  is the 3$\sigma$ exclusion obtained by varying $\leff$ in its 1$\sigma$ range \cite{Xenon100}. The blue dashed line comes from the conservative choice of having $\leff=0$ below 4 keV.

To determine the {\xenon} bounds we follow \cite{xenon10} doing an event-by-event fit to the three observed events at energies 6, 19 and 22 keV assuming a uniform background in the region $4\, {\rm PE}\leq {\rm S1}\leq 30\,{\rm PE}$ normalized to the total number of expected events, 1.8. We include finite resolution effects through a gaussian smearing of the S1 signal with a width given by $0.5\sqrt{S1}$. We include an S1 peak finding efficiency in the same way it is done in \cite{gondoloxenon}. We find this to be a small correction.

\subsection{XENON10}
The high {\sc Xenon100} ({\sc Xenon10}) threshold, roughly 8 keV (5 keV), is related to the fact that recoils of too small energy are not efficiently converted into a primary scintillation signal S1. It is through comparison of this signal to the secondary scintillation (S2 signal) that the background from electronic recoil is subtracted.

A low threshold analysis (1 keV) of {\sc Xenon10} data is available \cite{xenon10low} by discarding the S1 signal altogether and looking just at the S2 signal. Backgrounds are thus allowed to pollute the data but the gain in sensitivity at lower masses is substantial.

Since at low recoil energies the scintillation photons do not give a measurable S1 signal, one is forced to calibrate the energy scale using just the S2 signal. This calibration is encoded in the so called $\mathcal Q_y$ parameter defined as
\begin{equation}
\mathcal Q_y(E_{{R}})\equiv\frac{N_e(E_{{R}})}{E_{{R}}},
\end{equation}
where $N_e$ is the number of measured photoelectrons for a recoil of a given energy $E_{{R}}$. The role of $\mathcal Q_y$ is similar to the one of $\leff$. Fixing a lower threshold for the magnitude of the S2 signal will, depending on $\mathcal Q_y$, define a lower threshold on the nuclear recoil energies probed by the experiment: the smaller the $\mathcal Q_y$ the higher the effective threshold. Since the value of  $\mathcal Q_y$ is not measured below 4 keV \cite{Manzur:2009hp, Horn:2011wz} extrapolations must be used.

{\sc Xenon10} takes its S2 threshold at 5 photoelectrons, corresponding to roughly 1.4 keV with their choice of $\mathcal Q_y$. According to \cite{Collar:2011wq} this choice is far too generous. \cite{Collar:2011wq} thus proposes a smaller $\mathcal Q_y$ (see details in the original reference) corresponding to an energy threshold of roughly 4 keV.

To calculate the {\sc Xenon10} bounds we use the observed events in Fig.~3 of \cite{xenon10low}, corresponding to 15 kg\,days of effective exposure,  passing all the 5 cuts detailed in Tab.~1 therein. In a fashion which resemble the $p_{\rm max}$ method \cite{Yellin:2002xd}, we determine the two consecutive events between which the total expected signal from DM is maximized. We use poissonian statistic to define the likelihood,
\begin{equation}
\mathcal L_{\rm Xe10}=e^{-S2_{\rm DM}}\,,
\end{equation}
where S2 is the number of photoelectrons (including acceptances and poissonian statistical fluctuation which affects the S2 resolution) expected in the above interval under the DM hypothesis. We thus use $\chi^2_{\rm Xe10}=-2\ln\mathcal L_{\rm Xe 10}$ to set bounds. In the rest of the paper our choice for $\mathcal Q_y$ is the same as the one adopted by the {\sc Xenon10} collaboration. A comparison with the more conservative choice of  \cite{Collar:2011wq} is shown in Fig.~\ref{xenonbound}.

 \subsection{CDMS-Si}
The cryogenic CDMS experiment ( performed at the SUL like CDMS) operates Silicon ($A\approx28$, $Z=14$)  and Germanium solid-sate detectors. Like {\xenon}, CDMS has the ability to discriminate between nuclear recoils and electronic backgrounds, measuring both ionization and phonon signals. We use the results from the unofficial analysis \cite{filippinithesis, filippinitalk} of 6 Si detectors corresponding to a raw exposure of  53.5 kg\,d. We use the efficiency reported in \cite{filippinithesis, filippinitalk} which drop to zero below 7.8 keV. No events are observed with an expectation of 1.1 events from surface backgrounds. We use the poissonian likelihood,
\begin{equation}
\mathcal L_{\rm Si}=e^{-N_{\rm DM}}\,,
\end{equation}
where $N_{\rm DM}$ is the expected number of events under the DM hypothesis to extract the bounds.

As noted in $\cite{hooperdamacogent}$, CDMS-Si data indicate a behavior for the Silicon quenching factor which is not in accord with the prediction of Lindhard theory (see for instance Fig.~3.20 of \cite{filippinithesis}). This can be ameliorated by correcting the energy scale by $\mathcal O(20\%)$, which also goes in the direction of weakening the  CDMS-Si bounds but we do not include this correction in the fits.

 \subsection{CDMS-Ge}
For the analysis of CDMS Germanium we employ the data obtained in the recent low-energy analysis~\cite{Ahmed:2010wy}, with an energy threshold of 2 keV. Since the electronic and the nuclear recoil bands merge at these low energies, the sensitivity to low WIMP masses offered by this low-threshold analysis comes at the price of accepting a large amount of background. Though the CDMS collaboration provides possible explanations for the background and claims that it can explain all the events in the signal region, they neglect it as being dependent on extrapolation and on too many uncertainties. This approach has been recently criticized in various ways \cite{Collar:2011kf}.\\
In our analysis we follow the approach held by the CDMS collaboration not including backgrounds and treating, conservatively, all observed events as DM induced recoils.
We depart, for simplicity,  from the CDMS group retaining only the data coming from the best performing detector (T1Z5), that observed $36$ events between 2 and 20 keV. By considering all the events as signal we construct a $\chi^2$ by fitting the theoretical total rate $N_{\rm DM}$, so that
\begin{equation}\label{CDMSchi2}
\chi^2_{\rm Ge} = \frac{(N_{\rm DM}-36)^2}{36}\,\Theta(N_{\rm DM}-36).
\end{equation}
Our bound agrees with the one reported by the CDMS collaboration. Due to the large number of observed events (and in the absence of background subtraction), the  $\chi^2$ function in Eq. (\ref{CDMSchi2}) is rapidly growing thus giving confidence levels which are much closer to each other in comparison to the other experiments. This causes a deterioration of the global fit once the CDMS-Ge bound is included.

\section{Results}\label{res}

In this section we present the best fit results for the experiments discussed above.  Our approach here is to study the influence of different forms of cross-sections and velocity distributions, as discussed in the introduction, on the fits to the data.  This approach differs somewhat from the usual effective theory one, where the bounds on different operators are studied.  The virtue of the current method is that it allows one to identify the  physical necessities in order to minimize the experimental tension.   Realistic scenarios of course, often require taking  linear combinations in either approaches.

\subsection{Format of the figures}\label{plots}

In our figures, all continuous (dotted) curves correspond to $95\%$ ($99.7\%$) C.L.\ for two degrees of freedom,
i.e.\ $2\sigma$ ($3\sigma)$ corresponding to \ $\Delta \chi^2 = 6$ ($\Delta\chi^2 = 11.6$).  In the figures we show the following:
\begin{itemize}
\item In purple, the region favored by the modulation observed by DAMA at fixed value of $q_{\rm Na}$.
The parameter fixed by the fit is essentially $M_{\rm DM}\times q_{\rm Na}$, such that considering a higher $q_{\rm Na}$ value
shifts the favored
region to lower DM masses.  The text `DAMA' lies over the DAMA best fit point.
\item In yellow,  the wide region favored by the modulation observed by \cogent.  We further impose that the DM rate alone does not exceed the rate observed by  \cogent{} in any point.
%The `CM' text lies over the best-fit point.
In view of the poor statistical significance, we plot here the $68\%$ C.L. contour (dashed-dotted) as well as the $95\%$ C.L.\ contour.

\item In red, the favored region for the rate observed by \cogent{} (data in the right panel of fig.~\ref{cogentunmod}) assuming, in addition to the DM signal, an energy-independent constant term with normalization fixed as in Eq. (\ref{bestfitcog}).
This assumes, arbitrarily, that the L-shell decays of activated isotopes account for the majority of low energy background events.
The region favored by the \cogent{} rate up to this caveat is shown in red  and marked as `CR'.

\item A green curve for the bound obtained with the CDMS Silicon result, denoted as `Si'.

\item A red curve for the  bound by CDMS with Germanium, denoted as `Ge'.

\item A blue curve for the bound by \xenon, denoted as `Xe$_{100}$'.

\item A purple curve for the bound by {\sc Xenon10}, denoted as `Xe$_{10}$'.

\item A dark yellow curve for the bound from SIMPLE, denoted as `C Cl F'.

\item Finally, we plot a green dot indicating the global best fit taking into account all signals and bounds.
\end{itemize}
We define the global $\chi^2$ as $\chi^2 = \sum_i \chi^2_i$, summing over all the relevant experiments, and
report the value of $\chi^2$ for the best fit point.
Such $\chi^2$ can be used as statistical indicator to compare different fits and to evaluate the overall quality of the fit
(although other more sensitive statistical indicators exist).
For experiments that give bounds we fixed $\chi^2_i=0$ when no DM is present, such that we expect that a good fit should correspond to
$\chi^2\sim n_{\rm obs}-n_{\rm par}$, where $n_{\rm obs}$ is the number of observed data-points and
$n_{\rm par}$ the number of free parameters.
Since the {\cogent} rate could be contaminated by unknown backgrounds at low energy, we perform two global fits:
\begin{itemize}
\item[i)] Fitting all data including the {\cogent} rate.
A good fit should have $\chi^2_{\rm i)} \sim  43$, as we fit 35 data points in the {\cogent} rate, $9$ in the DAMA modulation, $2$ in the {\cogent} modulation with
a number of free parameters going from $2$ to $4$;
\item[ii)]  Fitting all data (the {\cogent} modulation and all other experiments) but dropping the {\cogent} rate.
A good fit should have $\chi^2_{\rm ii)} \sim 8$.
\end{itemize}
In the pictures we report the value of both $\chi^2$ evaluated at their best-fits, using the format $``\chi^2  =\chi^2_{\rm i)},\chi^2_{\rm ii)}$''.  Our Results are presented in the next sub-sections.

\begin{figure}[ht!]
$$\includegraphics[width=0.45\textwidth]{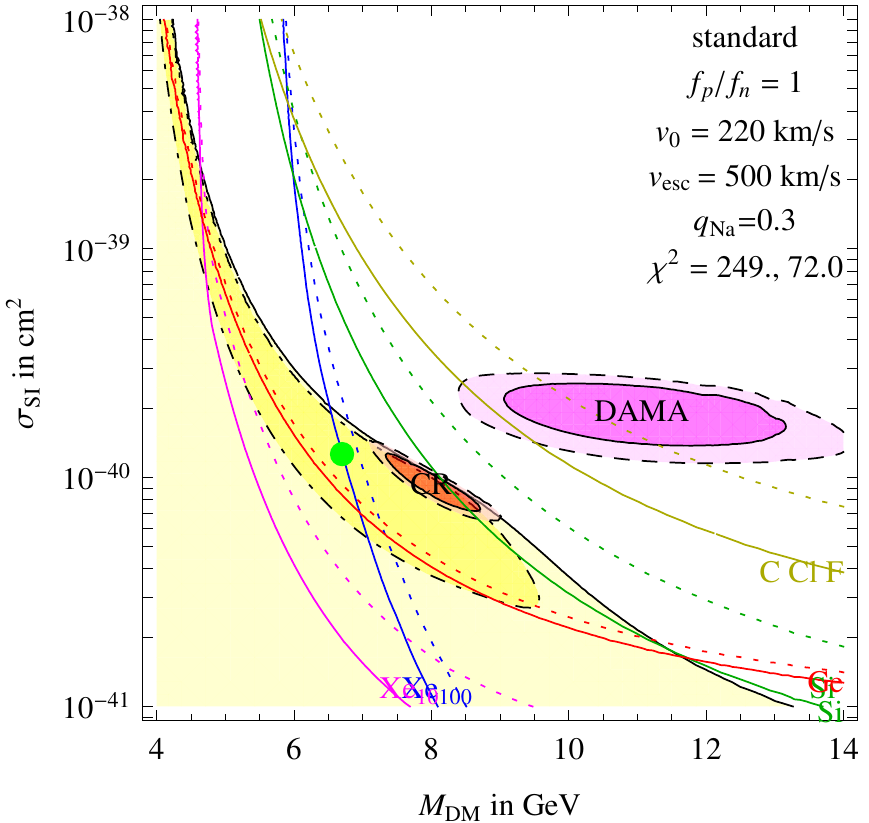}\qquad
\includegraphics[width=0.45\textwidth]{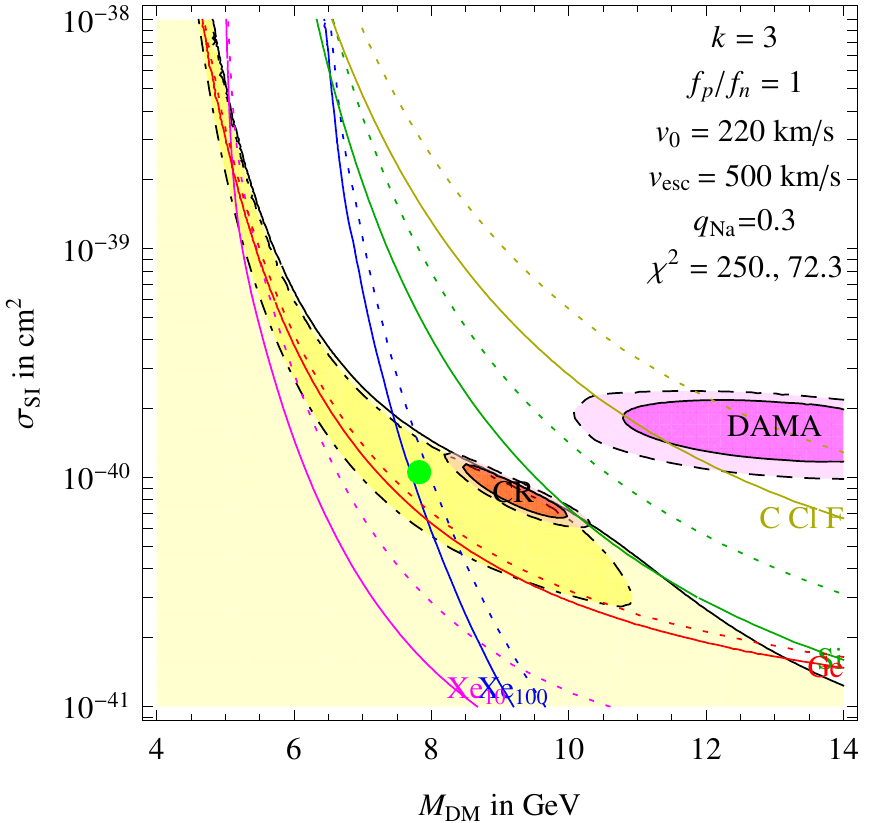}$$
$$\includegraphics[width=0.45\textwidth]{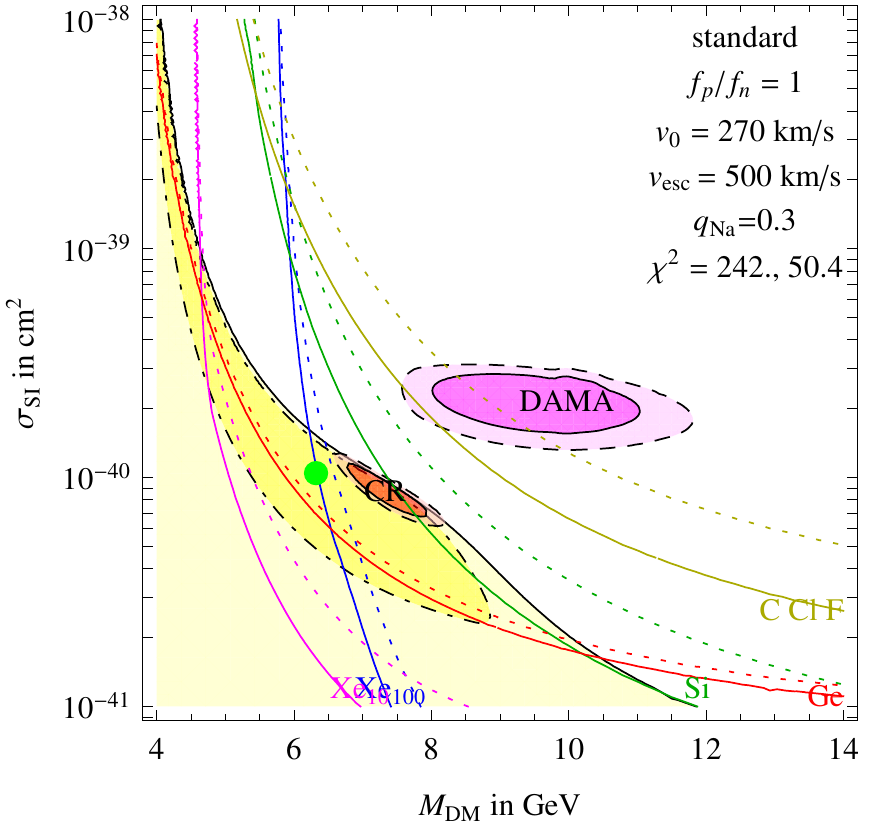}\qquad
\includegraphics[width=0.45\textwidth]{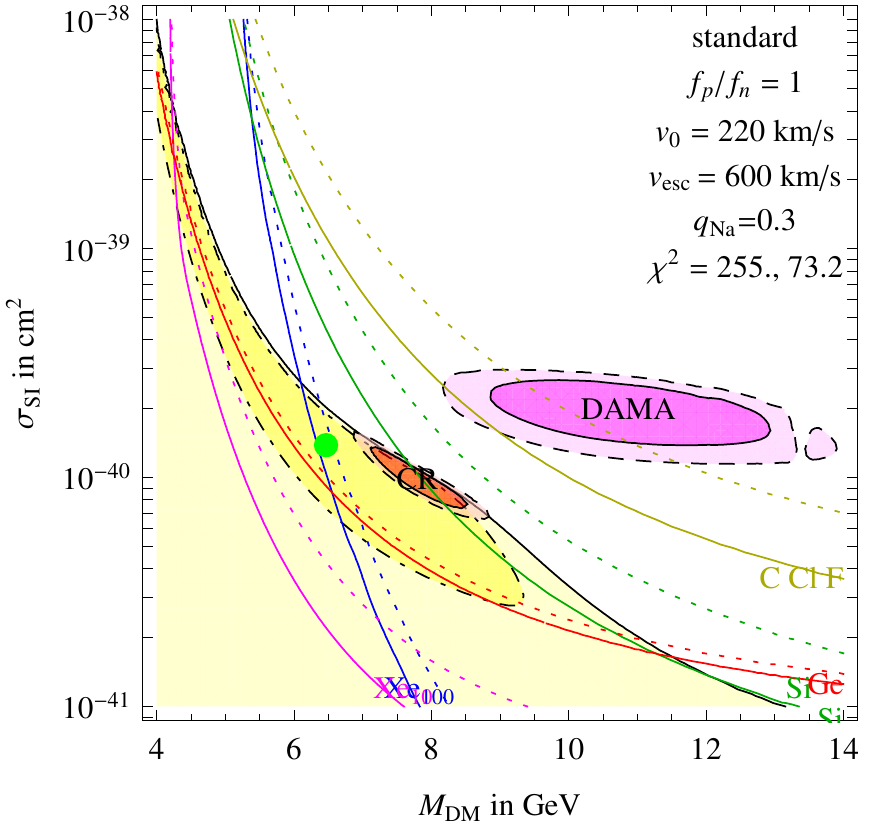}$$
\caption{\em {\bf Upper left:} Standard fit.  DAMA and \cogent{} do not overlap, and are excluded by many experiments.
In the other plots we vary the DM velocity distribution, finding minor changes.
On the {\bf top right:} we use the smooth distribution of Eq.~\eqref{eq:fk} with $k=3$.
{\bf Bottom left} A higher $v_0=270\, {\rm km/s}$ and lower $v_{\rm esc}=500\,{\rm km/s}$ are assumed. {\bf Bottom right:} A higher $v_{\rm esc} = 600\,{\rm km/s}$ is taken.  In all plots $f_p/f_n=1$ and $q_{N_a}=0.3$.
See Section~\ref{plots} for the color coding.% The color coding is described in% Sec.~\ref{plots}.
 \label{fig:standard}}
\end{figure}

\begin{figure}[ht!]
$$\includegraphics[width=0.45\textwidth]{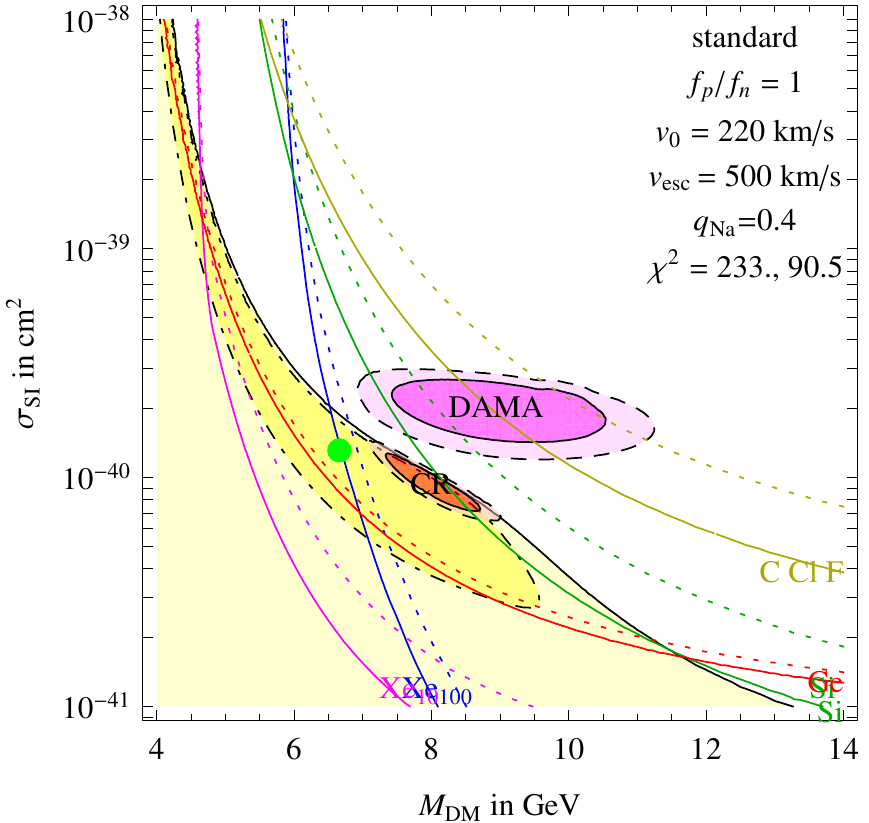}\qquad
\includegraphics[width=0.45\textwidth]{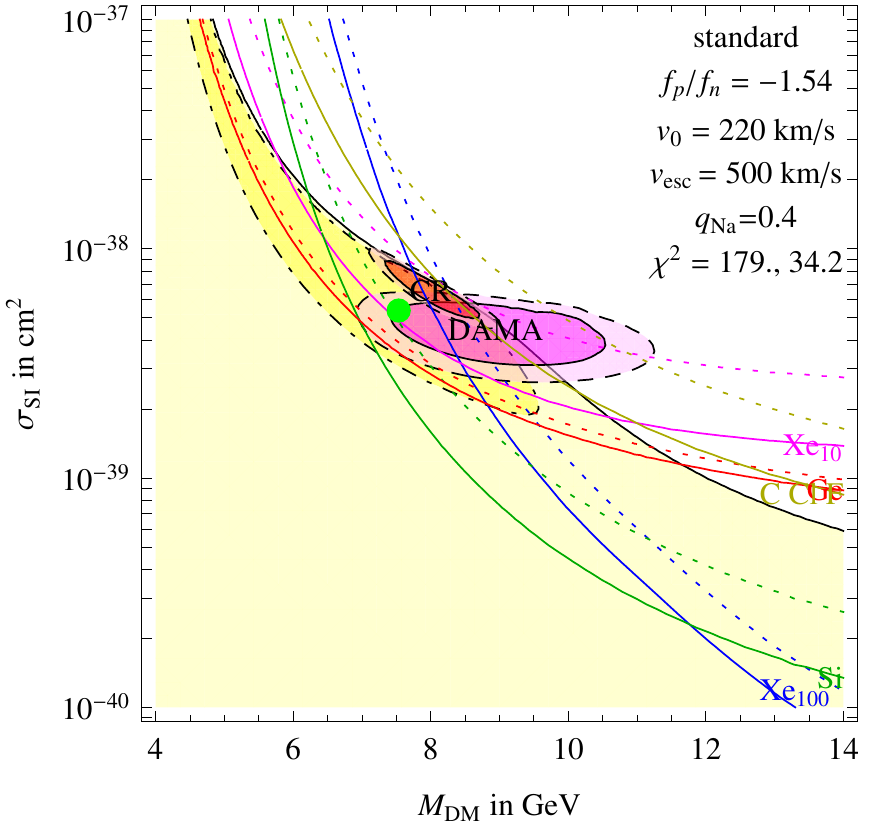}$$
$$\includegraphics[width=0.95\textwidth]{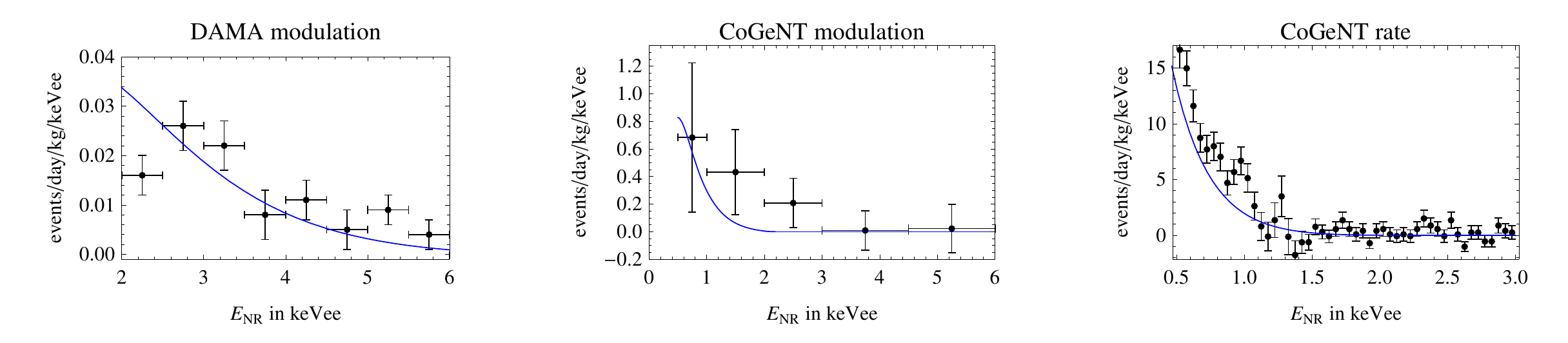}$$
\caption{\em {\bf Top left:} Standard fit  assuming a higher quenching factor for Sodium, $q_{\rm Na}=0.4$.
{\bf Top right:} Global fit assuming { isospin-violation}, dictated by the $f_p/f_n$ parameter.  The situation significantly improves, but the best fit (marked with a green dot) remains very poor.  Color coding is described in Section~\ref{plots}.
{\bf Bottom:} DM predictions for the best fit point, allowing a floating $f_p/f_n$.    The signal is plotted against the DAMA and \cogent\ modulated and unmodulated data.
\label{fig:standardf}}
\end{figure}

\subsection{Standard fit}
In Fig.\fig{standard}a  we show the ``standard'' fit, in terms of elastic spin-independent DM, using the cross-section in  Eq.~\eqref{eq:dsigmadE}, and assuming no form factor, $F_{\rm DM}=1$ and $f_n= f_p=1$.
We see that (i) DAMA and \cogent{} do not overlap and (ii) they are excluded or strongly disfavored by many experiments.
As a result the global best fit (green dot)
has a very high $\chi^2$, and corresponds to roughly no effect in DAMA.

\subsection{Astrophysical uncertainties}
\label{sec:astro}

We explore the  sensitivity to modifications in the velocity distributions (under the assumption of isotropy) by first considering the
smoothed cuts discussed in section~\ref{DMv}, as controlled by the parameter $k$:
bigger $k$ implies a smoother distribution while the sharply
cut, Maxwell Boltzmann distribution is obtained in the  $k\to 0$ limit.
These velocity distributions are shown in Fig.\fig{veldistr}a.
As can be seen in Fig.~\ref{fig:standard}b the fits to the experimental data
 assuming a DM velocity distribution with a smooth $k=3$ cut
are quite similar to the ones for $k\to 0$ (Fig.~\ref{fig:standard}a).
In view of the very minor difference from now on we  stick to sharp cuts and we
do not show the intermediate case $k=1$.

The improvement in changing $v_0$ is also small. The slope of DAMA (\cogent) spectrum fixes the value of $\mu^2 v_0^2/m_N$, hence for low-mass WIMPs raising $v_0$ favors smaller DM masses. The effects are shown in Fig.\fig{standard}c.
Similarly, changing the maximal DM velocity $v_{\rm esc}$ does not help, see Fig.\fig{standard}d.

\subsection{Quenching factor uncertainties in DAMA}
Assuming a quenching factor for Na higher than what is claimed by the DAMA collaboration, allows to shift the DAMA best fit region to lower $M_{\rm DM}$
(the combination $M_{\rm DM}\times q_{\rm Na}$ being essentially fixed), improving the fit.
In Fig.\fig{standardf}a we show the case $q_{\rm Na}=0.4$, that we will adopt from now on.
Even so, the best fit is so bad that we do not show its comparison with data.

\begin{figure}[t]
$$\includegraphics[height=5cm]{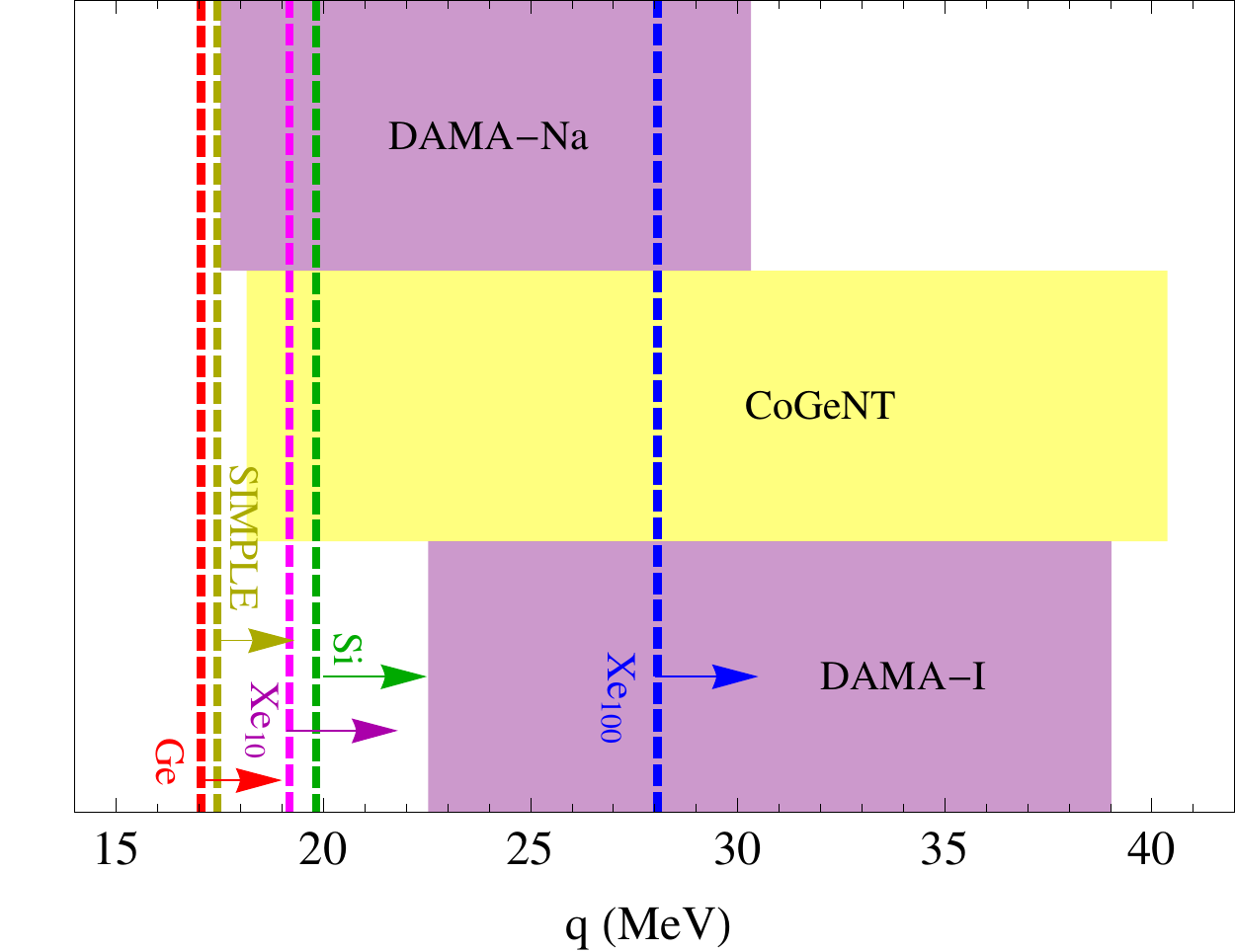}\qquad \includegraphics[height=5cm]{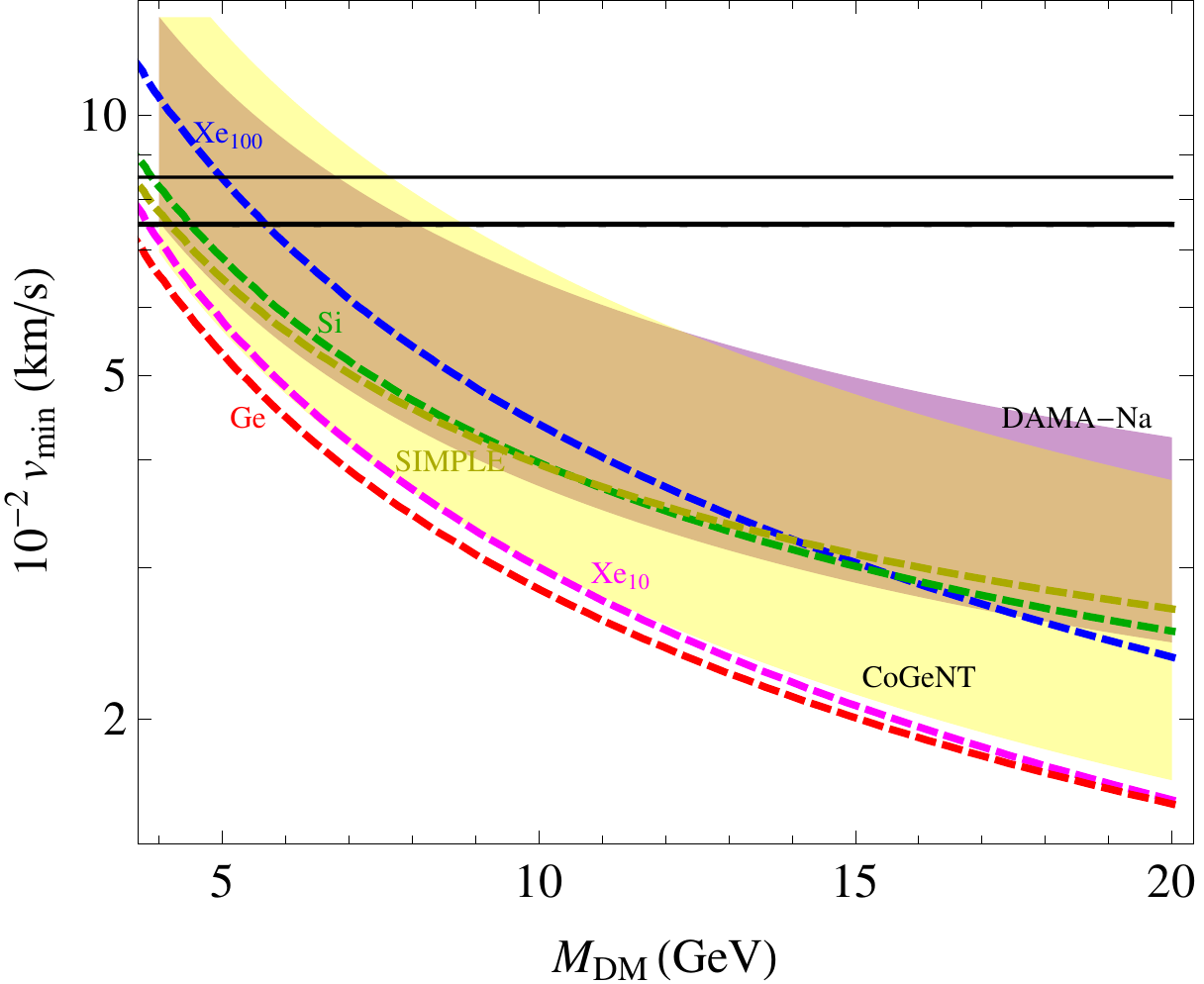}$$
\caption{\em Ranges of transferred momentum $q$ {\bf (left)} and minimal DM velocities, $v_{\rm min}$,  {\bf (right)}
probed by relevant experiments.
 Solid bands: {\cogent} 0.5-3.0 keVee  band (yellow) and DAMA  2-6 keVee  bands (purple) for both unchanneled Sodium (upper panel) and channeled Iodine (lower panel). Dashed lines: {\xenon}  3 keV  threshold (blue), CDMS-Si   7 keV  threshold (green), CDMS-Ge 2 keV threshold (red) and SIMPLE 8 keV threshold (dark yellow). The horizontal lines show the maximum possible DM velocity in the Earth frame for $v_{\rm esc}=500$ km/s (thick) and $v_{\rm esc}=600$ km/s (thin).
\label{fig:vmin}
\label{fig:qv}}
\end{figure}

\subsection{Isospin violating couplings}\label{fpfn}
It is possible that DM does not couple equally to protons and neutrons.  Such isospin-violating DM scattering has been studied against the \cogent\ rate data previously in~\cite{Chang:2010yk,Kang:2010mh,Feng:2011vu}.   In Fig.\fig{standardf}b (and in the rest of the fits) we relax the assumption of $f_p=f_n$, and  allow for different spin-independent cross sections on protons and neutrons.
As can be seen, floating the  additional  parameter, $f_p/f_n$, allows to
\begin{itemize}
\item Improve the agreement between \cogent{} and DAMA.
\item Ameliorate the tension of the positive results with one experimental bound, through the tuning of $f_p Z+f_n(A-Z)\approx 0$.  In our fits, we take into account the different isotopes in the various experiments, thereby allowing for only a partial cancelation.
\end{itemize}
When performing a global fit, we find the best-fit shown in Fig.\fig{standardf}b, with a significant improvement over the $f_p=f_n$ case
 shown in Fig.\fig{standardf}a.
Still, the global fit remains very poor, despite the assumption of a relatively large Sodium quenching factor, $q_{\rm Na}=0.4$.
 Indeed,  the incompatibility of the tentative DM signals is with four experiments, performed with three different nuclei (Xe, Si and Ge).  Consequently, this extra parameter allows to avoid the strongest constraint (coming from Xe experiments), but
 is insufficient to relieve the tension completely.

\begin{figure}[t]
$$\includegraphics[width=0.45\textwidth]{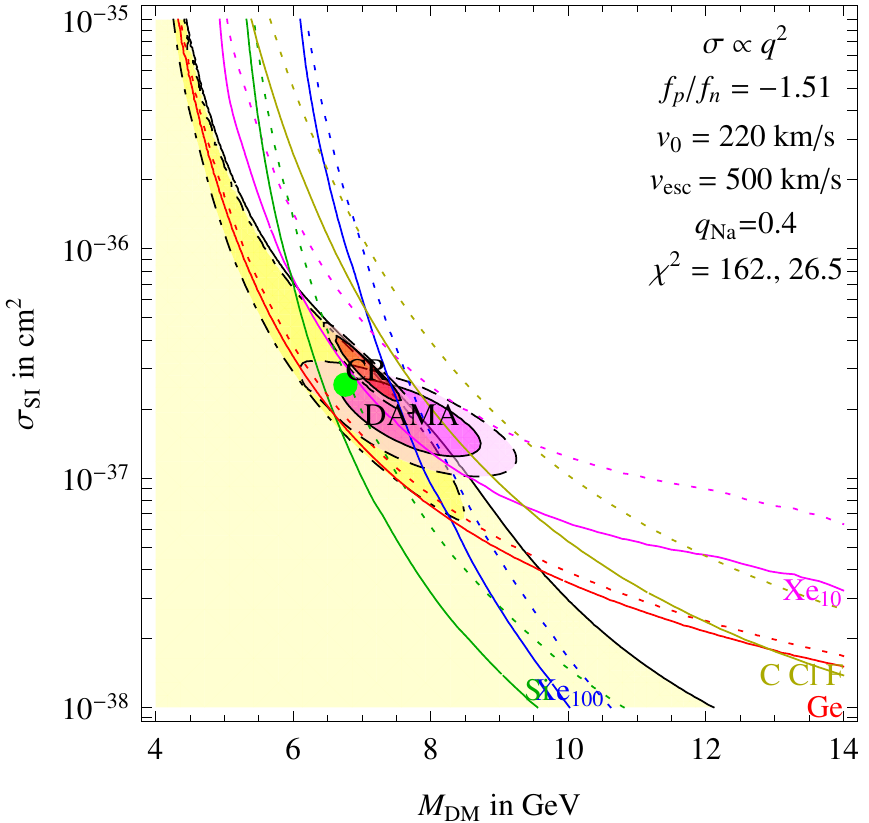}\qquad\includegraphics[width=0.45\textwidth]{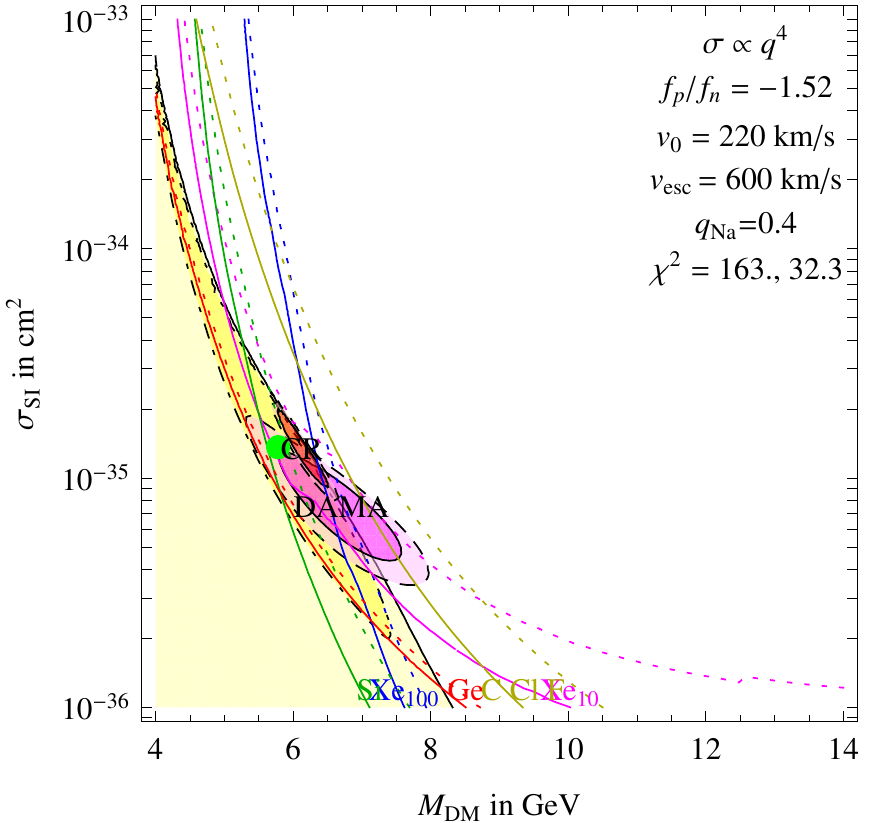}$$
$$\includegraphics[width=0.45\textwidth]{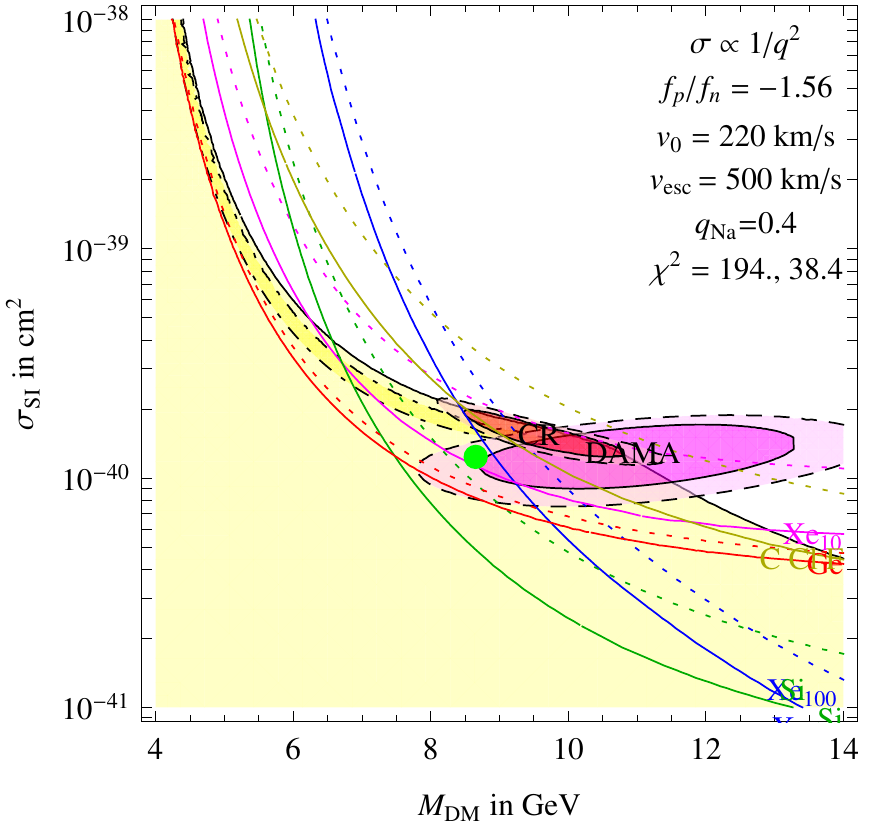}\qquad\raisebox{3cm}{
\parbox{0.45\textwidth}{
\caption{\em Fit with free $f_p/f_n$ and with { momentum-dependent form factor}.
In the left upper (lower) row the cross section grows (decreases) with  transferred momentum $q^2$.  The top right plot shows the fit for a $q^4$ form factor.
Color coding is given in section~\ref{plots}.
\label{fig:standardq}}}}$$
\end{figure}
%
%\begin{figure}
%$$\includegraphics[width=0.45\textwidth]{figs/fitstandardfqm2}\qquad
%%\includegraphics[width=0.45\textwidth]{figs/fitstandardfq4}
%$$
%\caption{\em Fit with free $f_p/f_n$ and with cross section that decrease with transferred momentum $q^2$.
%\label{fig:standardqm}}
%\end{figure}

\subsection{Momentum-dependent elastic scattering}
Momentum dependent scattering arise in several instances.  One notable case is when the DM-nucleon interaction is mediated by a pseudo-scalar.  It is possible to systematically study the momentum dependent effects by utilizing non-relativistic effective theory.  Indeed, the typical energy transfer in DM-nucleus collisions relevant for direct-detection is much below the  nuclear binding energy.  Additionally, the DM velocity, $v\simeq 10^{-3}$, is  non-relativistic.
%, the momentum transfer $q\approx A^{1/2}\,{\rm MeV}$ is much smaller than the typical inverse size of the nucleus $\approx A^{1/3}\,{\rm GeV}$. Furthermore the process is non-relativistic due to the small value of the DM-nucleus relative velocity $v/c\approx 10^{-3}$.
Consequently, it is possible to describe the scatterings via an  effective theory assuming a rotationally invariant potential \cite{Fan:2010gt}. Restricting  to spin-independent interactions, one finds to leading order in the DM velocity, $v$, and the momentum transfer, $q$, only four  scalar operators from which the effective potential may be constructed.   Written in momentum space they are,
\beq \begin{array}{llll}\label{sys:factors}
\mathcal A_1\propto 1, &&& \sigma \propto 1\,,\\
\mathcal A_2\propto \vec s_{\rm DM}\cdot \vec q,&&& \sigma \propto q^2\,,\\
\mathcal A_3\propto\vec s_{\rm DM}\cdot \vec v,&&& \sigma \propto v^2\,,\\
\mathcal A_4 \propto \vec s\cdot\vec q\times \vec v,&&& \sigma \propto q^2 v^2\,,
\end{array}\eeq
where $\vec s_{\rm DM}$ the spin of the DM.
In a Lorentz invariant theory, specific linear combinations of the $\mathcal A$s  appear in the full ${\rm DM}\, N\to {\rm DM}\,N$ amplitude.
%A specific example, for the case of magnetic coupling of DM to photons, will be studied in section~\ref{sec:magnetic}.
For further discussion, see \cite{Fan:2010gt}.

From a phenomenological point of view we can study the effect of $v$- and $q$-dependent form factor on the compatibility between DAMA, \cogent{} and the null-experiments.
We consider the following DM form factors that enter in Eq.~\eqref{eq:dsigmadE}:
\begin{equation}
F^2_{\rm DM} = q^2/q_{\rm ref}^2~, \quad
q^4/q_{\rm ref}^4~,\quad
q_{\rm ref}^{2}/q^2~,\quad
q_{\rm ref}^{4}/q^4\,,
%\frac{d R}{d E_R}=N_T\frac{\rho_\odot}{M_{\rm DM}}\int_{|\vec v|>v_{\min}} d^3v\,v\left(\frac{v}{c}\right)^{n_v}\,f_\odot(v,t)\,\left(\frac{q}{q_{\rm ref}}\right)^{n_q}\frac{d\sigma_{{\rm DM}-N}}{d E_R}
\end{equation}
 where the momentum transfer is normalized with respect to the reference $q_{\rm ref}=100$\,MeV.
 The last form factor can be generated in models where DM-nucleon scattering is mediated by a new particle with mass
 $m\ll q_{\rm ref }$.  The $1/q^2$ form factor can arise in this model with a coupling of type ${\cal A}_2$.

A momentum-dependent form factor changes the relative rate in different experiments according to the range of $q^2$ they probe, as summarized in Fig.\fig{qv}a.
We show bands corresponding to the signal regions for DAMA (2-6 keVee) and for {\cogent} (0.5-3 keVee), as well as lower $q^2$ value corresponding to the energy threshold for all other experiments. For SIMPLE we show the $q^2$  corresponding to fluorine which is target element with the largest rate thus being the main contributor to bounds.

$q^2$ and $q^4$ form factors help in this respect since they deplete the spectrum at low recoil energies and hence require smaller masses for a good fit. The \xenon\, bounds are expected to get stronger relative to the other experiment due to the higher momentum transfer. It must be noted, however, that in the low mass region, the sensitivity of \xenon\, is saturated by its threshold (and consequently the bound becomes vertical at small masses).
Another feature of $q^n$ ($n>0$) form factors is that they enhance the {\cogent} signal with respect to the DAMA one, since the former probes a higher $q$ region. This implies that the ratio of the best-fit cross sections for {\cogent} and DAMA will be smaller with respect to the standard fit, making the separation of the two regions more severe. This problem can be ameliorated by an appropriate choice of $f_n\neq f_p$.
Fig.\fig{standardq}a shows the global fit.  We find that $q^2$ form factor gives mildly better fits than the standard $q^0$ case, however the overall $\chi^2$ remains poor.  Similar results are found for a $q^4$ form factor.

An analog, but inverted, discussion apply for $1/q^{2}$ or $1/q^4$ form factors.
These form factors  move   both the  DAMA and {\cogent} regions to higher masses and  make the two best-fit points closer (assuming $f_n=f_p$). The DAMA non-modulated signal, which probes recoil energies below the ones where modulation is observed, excludes the $q^{-4}$ behavior.
Fig.\fig{standardq}b shows the results for $1/q^2$.  We do not find that the  form factor improves the global fit.

\begin{figure}
$$\includegraphics[width=0.45\textwidth]{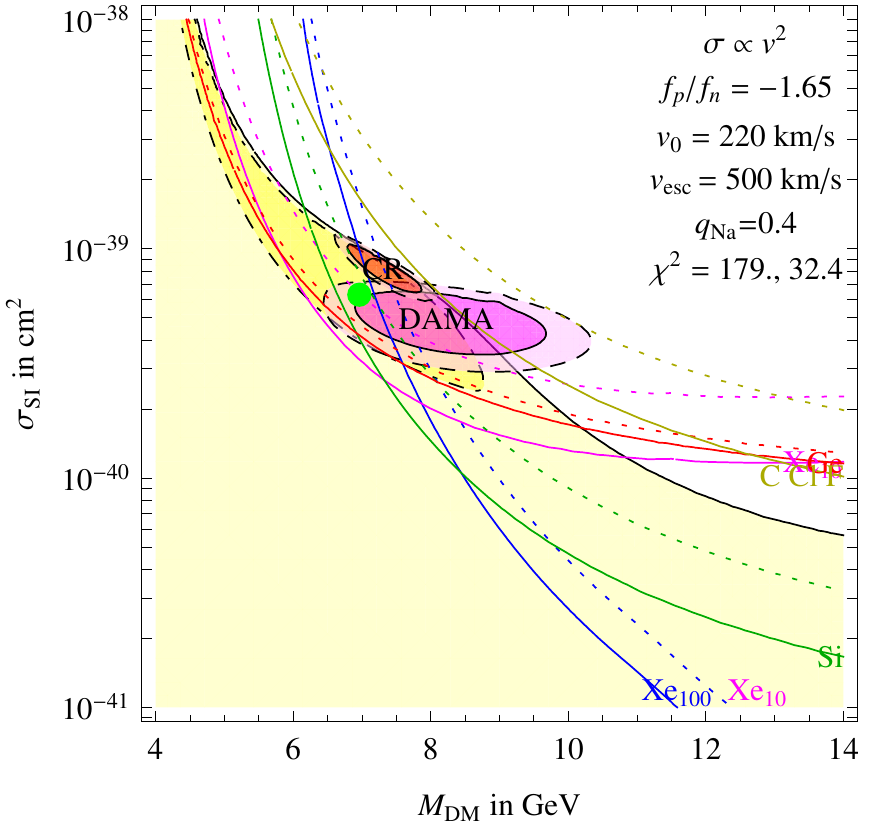}\qquad
\includegraphics[width=0.45\textwidth]{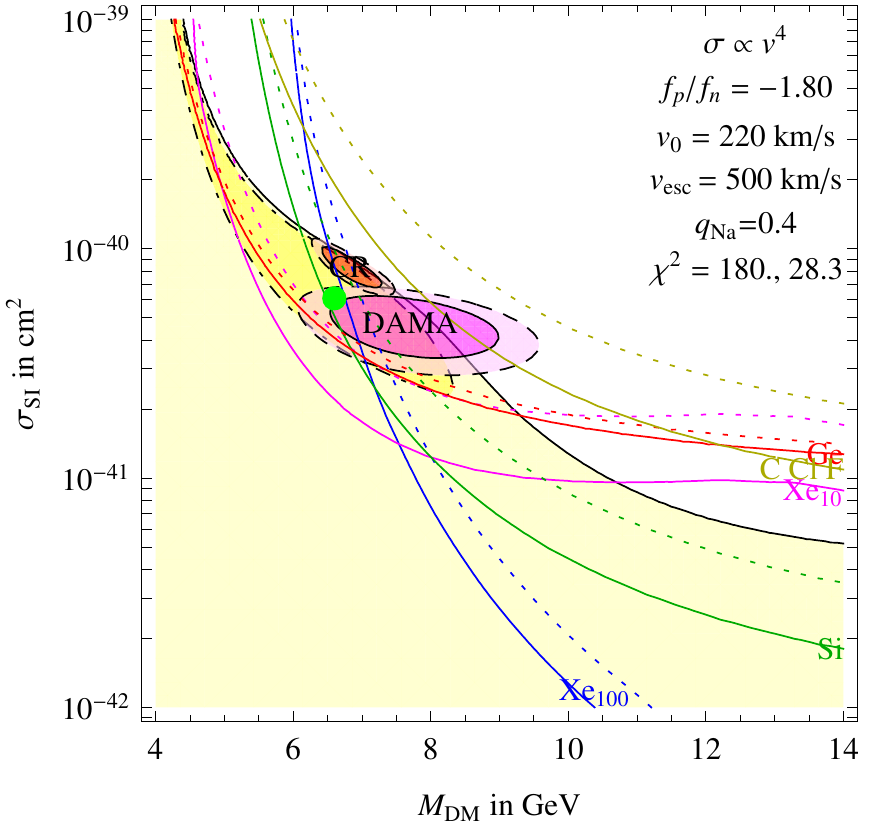}$$
\caption{\em Fit with free $f_p/f_n$ and with a velocity dependent form factor.  {\bf Left:} $v^2$ form factor.  {\bf Right:} $v^4$ form factor.  An overall normalization effect is visible, as discussed in Section~\ref{DMv}.  For further discussion see Section~\ref{sec:velocity}.  Color coding is given in section~\ref{plots}.
\label{fig:standardv}}
\end{figure}

%%%%%%%%%%%%%%%%%%%%%%%%%%
\subsection{Velocity-dependent elastic scattering}
\label{sec:velocity}
%%%%%%%%%%%%%%%%%%%%%%%%%%%

The amplitudes in Eq.~(\ref{sys:factors}) also induce velocity-dependent form factors.   Here we take again the phenomenological approach and study the effect of the form-factors,
\begin{equation}
F^2_{\rm DM} = v^2/v_{0}^2, \qquad
F^2_{\rm DM} = v^4/v_{0}^4\,,
%\frac{d R}{d E_R}=N_T\frac{\rho_\odot}{M_{\rm DM}}\int_{|\vec v|>v_{\min}} d^3v\,v\left(\frac{v}{c}\right)^{n_v}\,f_\odot(v,t)\,\left(\frac{q}{q_{\rm ref}}\right)^{n_q}\frac{d\sigma_{{\rm DM}-N}}{d E_R}
\end{equation}
 where we normalize with respect to a mean velocity $v_0$.

The effect of the overall normalization discussed in Section~\ref{DMv} and shown in Fig.~\ref{fig:eta} is trivial, as can be seen in the fits displayed in Fig.~\ref{fig:standardv}.  To understand the effect of the sensitivity to the minimal velocity, we show the minimal DM velocity probed by different experiments  as a function of the  DM mass in Fig.\fig{vmin}b. We note that the minimal velocity does not depend on astrophysics (i.e.\ velocity distributions) nor on the scattering cross sections (see also~\cite{Fox:2010bz}).
The bands for {\cogent} and DAMA corresponding to the observed signal energy range are shown.
%That is to say 0.5-3.0 keVee for {\cogent} and 2-6 keVee for DAMA, for both unchanneled Sodium (left panel) and channeled Iodine (right panel) as we are interested in light DM.
For null-experiments we  show the $v_{\rm min}$  corresponding to the respective energy threshold, where we have taken 3 keV as a representative value for {\xenon}.
For completeness,
the horizontal lines show the maximum velocity that a DM can have with respect to the Earth, given $v_{\rm esc}=500$ or 600 km/s.
We see that neither {\cogent} not DAMA probe values of $v$ inaccessible to other experiments.

Nonetheless, some improvement in the global fit occurs when the typical minimal velocity is around the tail of the $\eta_n$ function.  In that region, a small change in sensitivity to $v_{\rm min}$ results in a large change in the scattering rate.  Indeed our global fits shown in Fig.\fig{standardv} demonstrate a small improvement in the fits.

%The SIMPLE line is the result of the lowest $v_{min}$ among those probed by the three different target elements.  FALSE!??

%For completeness we also show the lines corresponding to the maximum possible velocity a dark matter particle can be seen with in the Earth frame

\begin{figure}[t]
$$\includegraphics[width=0.45\textwidth]{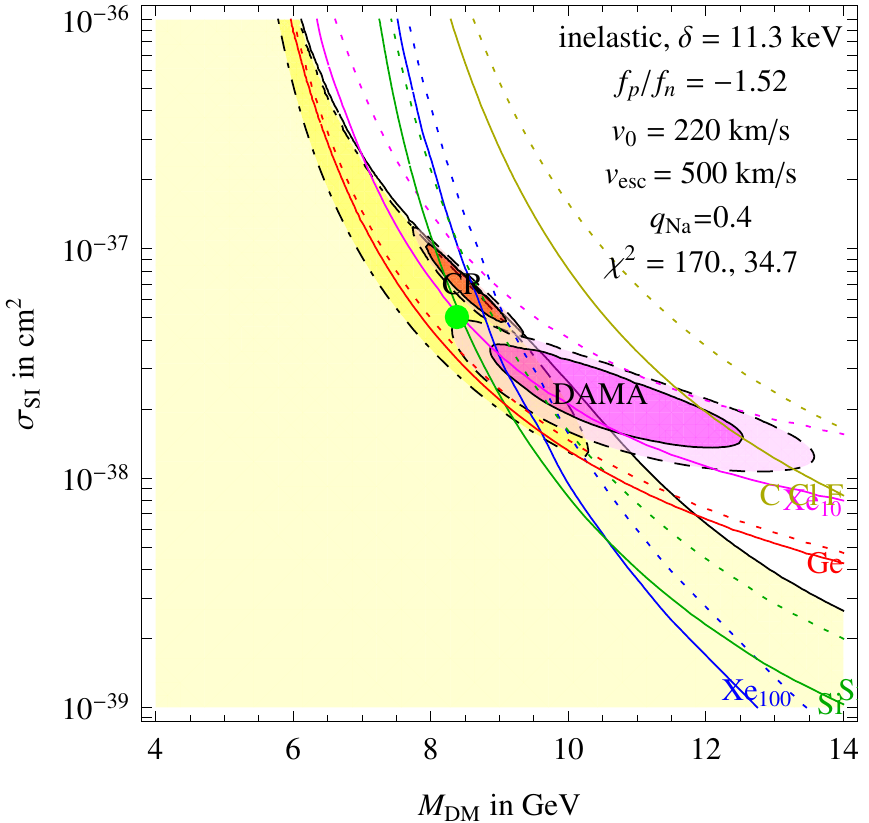}\qquad\includegraphics[width=0.45\textwidth]{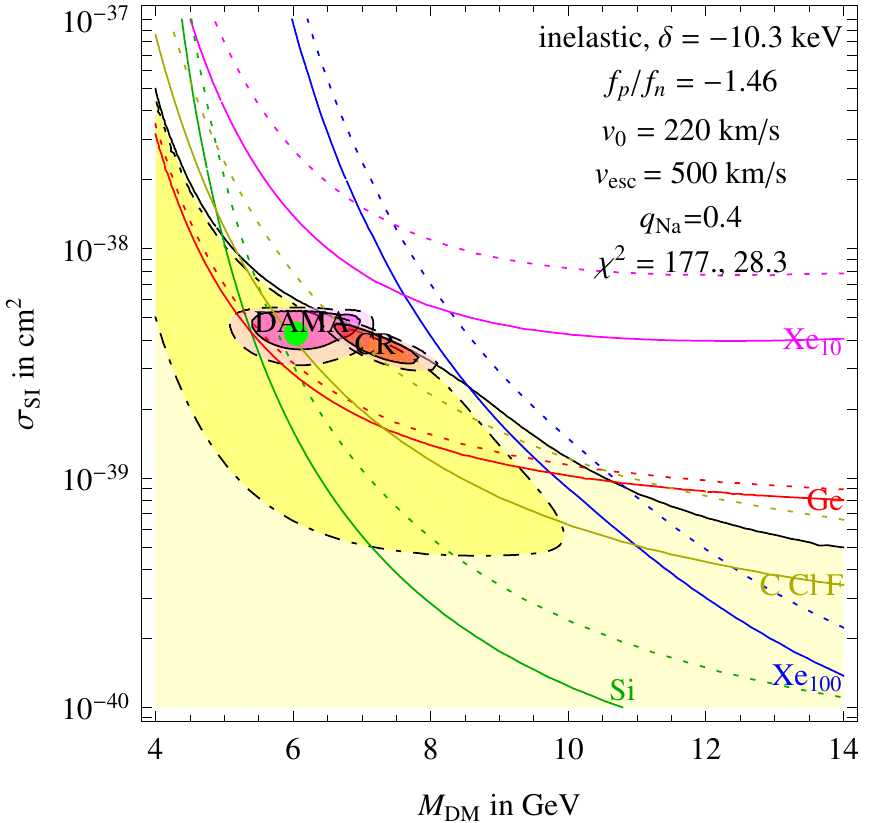}$$
$$\includegraphics[width=0.95\textwidth]{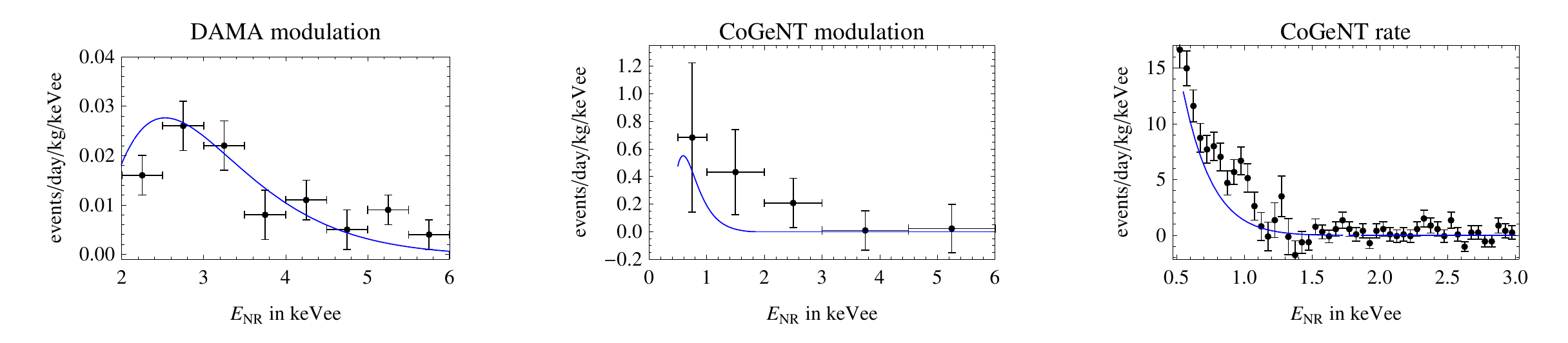}$$
\caption{\em Fit with {inelastic dark matter} and free $f_p/f_n$ and DM splitting, $\delta$.
{\bf Left:} An example of endothermic DM ($\delta >0$), which improve the global fit.
{\bf Right:} exothermic DM ($\delta<0$) improves the partial fit. Color coding is described in Section~\ref{plots}.
 {\bf Bottom:} DM predictions at the best fit point in the right panel (exothermic DM), plotted against the DAMA and \cogent\ modulated and unmodulated data.
\label{fig:iDM}}
\end{figure}

\subsection{Inelastic light Dark Matter scattering}
\label{sec:iDM}

It is possible that DM scatters inelastically  with the nucleus.   For inelastic scattering to take place, two semi-degenerate DM states are assumed, with mass splitting $\delta =M'_{\rm DM} - M_{\rm DM}$.
Up-scattering of the lighter DM state  requires it to have enough energy, thereby suppressing the rate for small values of the recoil energy. Since up-scattered iDM kinematically favors heavy targets, it was originally able to ameliorate the tension between the DAMA modulation and the null CDMS result~\cite{TuckerSmith:2001hy}, requiring $M_{\rm DM}\sim 100\,\GeV$ and $\delta\sim
100\,{\rm keV}$.

This original iDM scenario seems now excluded by the recent \xenon\, results~\cite{Aprile:2011ts,Farina:2011bh}. Here we study a different iDM regime with significantly smaller splitting and lighter DM (for previous study see~\cite{Chang:2010yk}).
In this window the tension with the null experiments is ameliorated mostly due to the small DM mass, while the DM scattering rate is falling above a few keV, much like in the elastic case.   The effect of $\delta$, is to modify the minimal velocity needed for scattering to occur, in accord with Eq.~\eqref{vminidm}.

For the masses and recoil energies of interest, the splitting $\delta$ is required to be smaller than $\sim 15$ keV in order to comply with the {\cogent} data.  Consequently, the inelasticity for up-scattering is only relevant for the light Sodium and Silicon targets in DAMA and CDMS respectively.
We find that such scenario does improve the global fit only slightly, and one example for $\delta \approx 11 $ keV is shown in Fig.\fig{iDM}a

It is possible, however, that the heavier DM state is cosmologically long lived, in which case it occupies a sizable fraction of the DM density.  In fact, this possibility occurs quite naturally, as noted in~\cite{Finkbeiner:2009mi,Essig:2010ye}.  In such a case, DM can also down-scatter with the nucleus, producing an exothermic reaction.  This possibility, dubbed exothermic DM (exoDM), was studied in~\cite{Essig:2010ye}. Technically, the rate for exoDM is given by the same expression as for the up-scattered iDM case, with now $\delta<0$. But the behavior of exoDM is significantly different.   The minimal velocity, Eq.~(\ref{vminidm}), is minimized for $E_R \simeq |\delta|\mu/m_N$ and hence lighter targets are more sensitive to exothermic DM scatterings. No net modulation in the total number of signal events is expected. However, since the spread of the spectrum around its maximum depends on the DM kinetic energy, the recoil energy spectrum will actually modulate annually. The modulated spectrum observed by DAMA can thus be reproduced~\cite{Essig:2010ye}.

In Fig.\fig{iDM}b we show the best fit for exoDM.  In order to isolate the effect of down-scattering, here we only consider down-scattered events, even though up-scattering is expected to be significant.
For the fit, we float the DM mass, the cross section, $f_p/f_n$ and the DM mass splitting, finding that
a $\delta \approx -10$ keV  improves the partial fit with respect to
the analogous fit with $\delta=0$, shown in Fig.\fig{standardf}b.

\begin{figure}[t]
$$\includegraphics[width=0.45\textwidth]{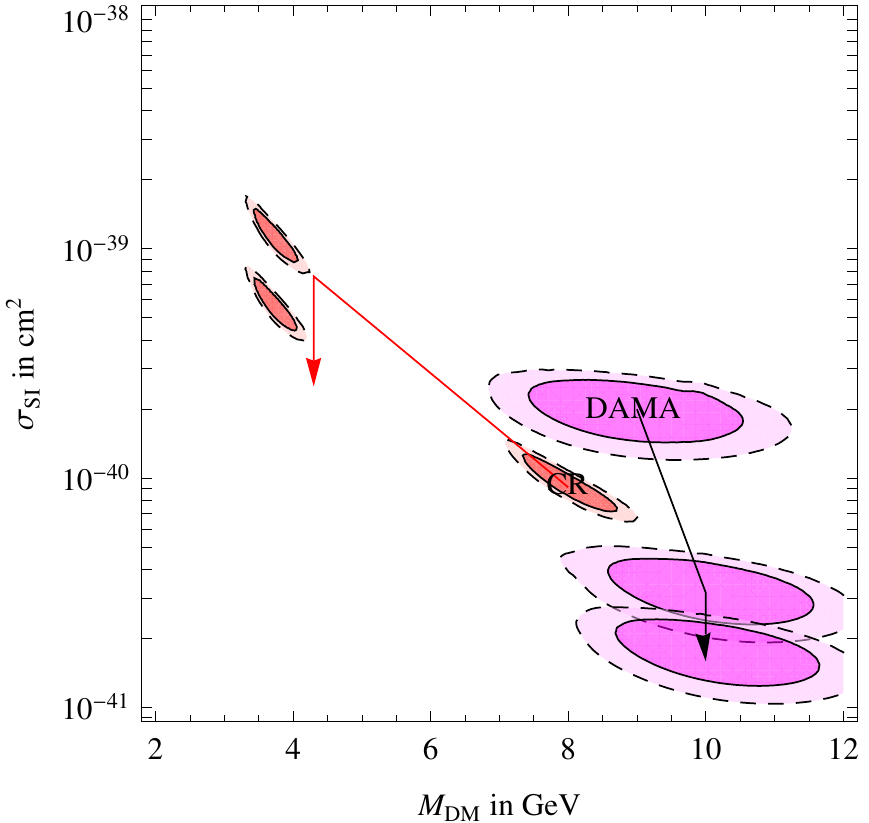}\qquad\includegraphics[width=0.45\textwidth]{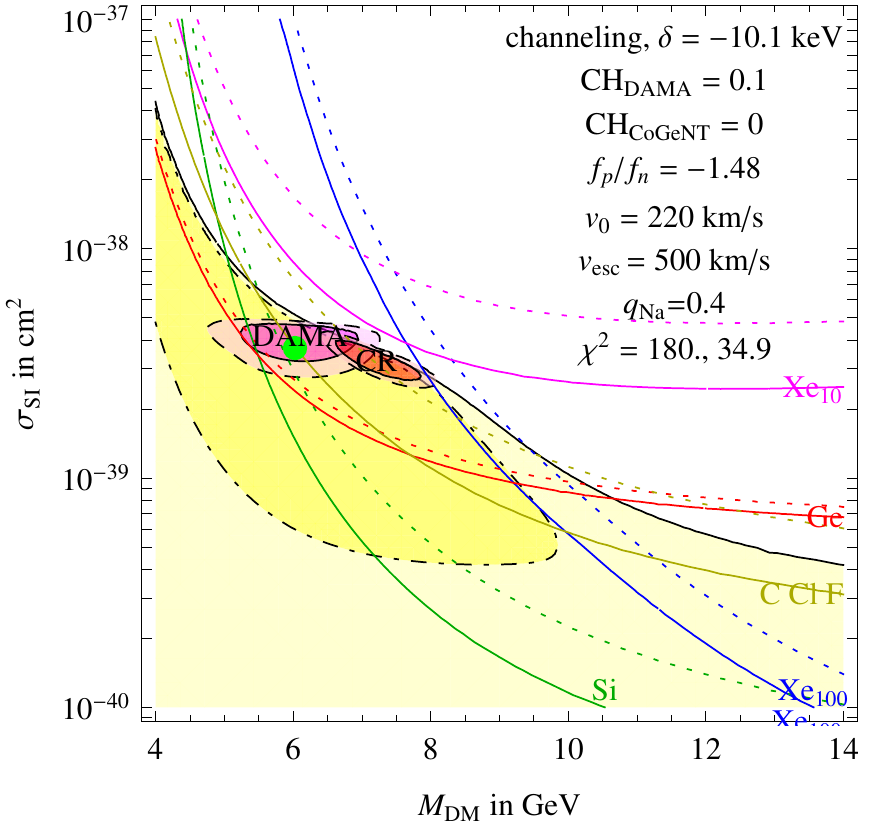}$$
$$\includegraphics[width=0.45\textwidth]{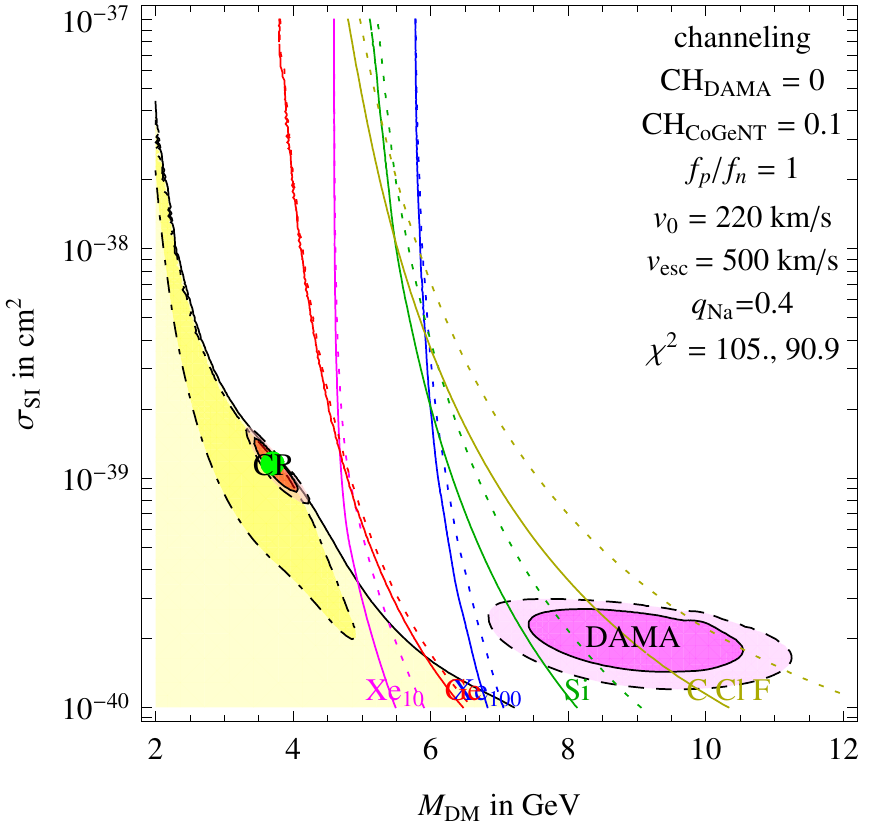}\qquad\raisebox{4cm}{
\parbox{0.45\textwidth}{
\caption{\em {\bf Upper left:}  regions favored by DAMA (in magenta) and  {\cogent} (in red) for $\delta=0$ and $f_p=f_n$
shift when increasing the channeling fraction $0\%\to 10\%\to 20\%$, as indicated by the arrows.
{\bf Upper right}:  fit with inelastic dark matter and free $f_p/f_n$ as in Fig.~\ref{fig:iDM}, but adding a $10\%$ channeling in DAMA.
 {\bf Bottom left}: channeling in {\cogent} alone. As can be seen, $10\%$ channeling allows for \cogent\ to evade all bounds while DAMA is still disfavored.  Color coding is described in Section~\ref{plots}.
\label{fig:CH}}}}
$$
\end{figure}

\begin{figure}[t]
$$\includegraphics[width=0.45\textwidth]{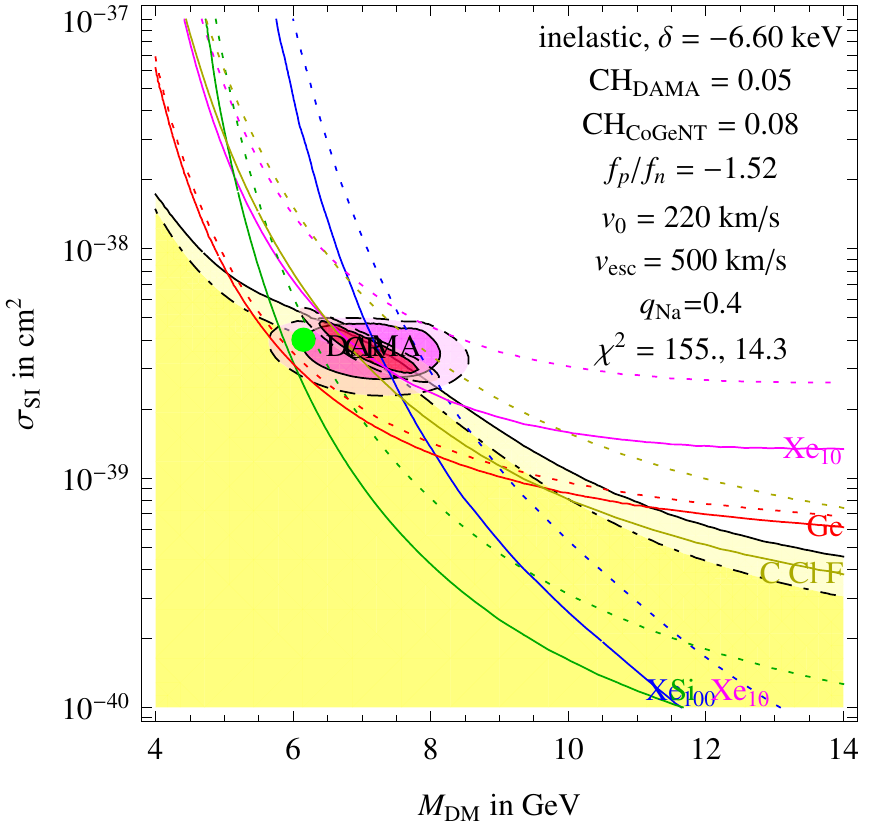}\qquad\includegraphics[width=0.45\textwidth]{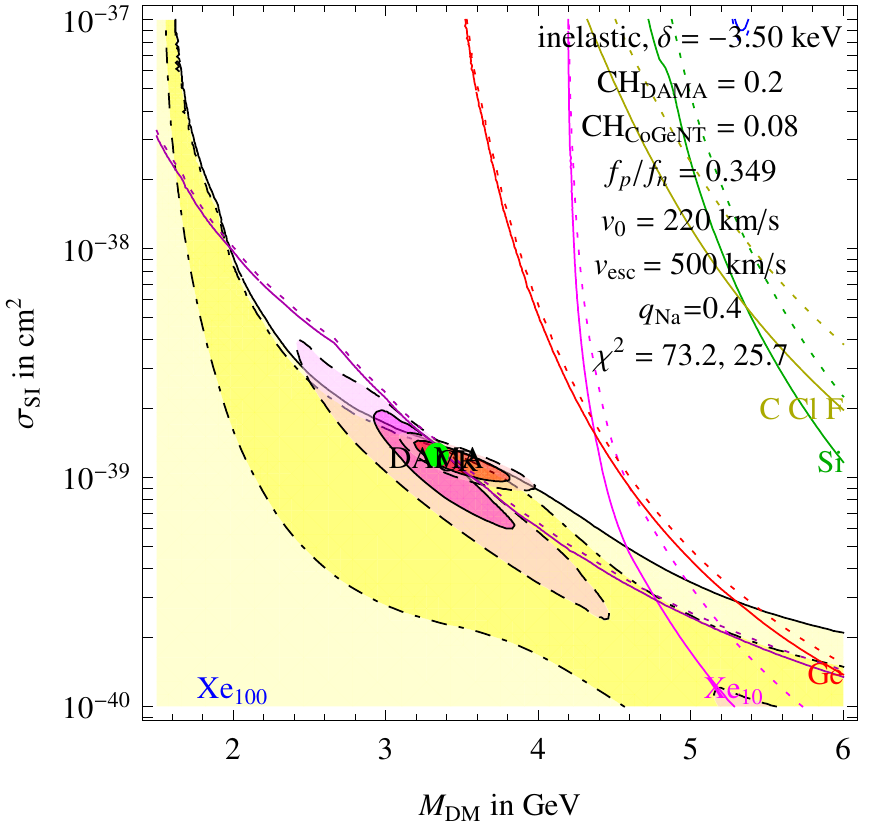}$$
$$\includegraphics[width=0.95\textwidth]{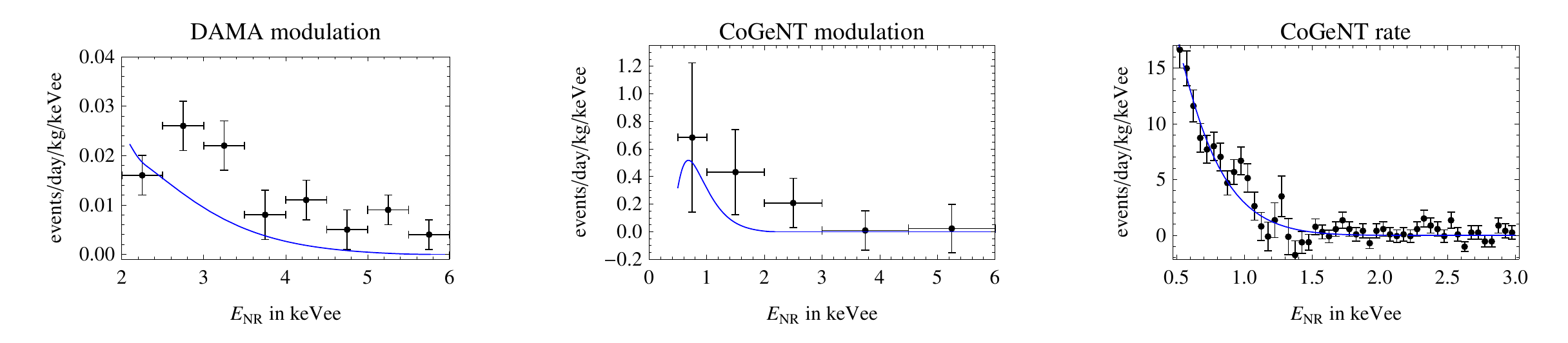}$$
\caption{\em Fit with channeling in both DAMA and {\cogent} together with inelasticity.
{\bf Left:}  global best fit at larger mass.
{\bf Right}:  new best fit at smaller mass as discussed in Section~\ref{sec:CHinelastic}. The dark magenta line is the bound from DAMA unmodulated signal.
 {\bf Bottom:} DM predictions at the best fit point in the right panel.
\label{fig:CH2}}
\end{figure}

\subsection{Channeling}
\label{sec:channeling}
We now allow  for some amount of channeling in the NaI(Tl) crystals of DAMA and/or in {\cogent}.
In this section we do not introduce any inelasticity: $\delta =0$.
Fig.\fig{CH}a shows how a relatively small fraction of channeled events is
enough to shift the best fit regions for these experiments to a different global minimum of the $\chi^2$.
Next, increasing the channeling fraction further, mildly shifts the best-fit regions down to lower values of the cross section, but
no new best-fit regions appear.
The effect of channeling can easily be understood:
\begin{itemize}
\item In {\cogent}, the channeled best-fit region has a DM mass reduced by a factor $\approx 0.5$ (the square root of the quenching factor in eq.\eq{qGe}).
\item
In DAMA, scattering on lighter Na (on heavier Iodine) dominates if channeling is negligible (significant), and the best-fit region
almost corresponds to the same DM mass, as shown in Fig.\fig{CH}a.
\end{itemize}
Adding channeling only in DAMA allows to get a best fit comparable to the un-channeled best fit.
Fig.\fig{CH}b shows one example of this possibility, assuming a $10\%$ channeling in DAMA.

Adding channeling also in {\cogent} moves its best fit to a smaller DM mass.
Interestingly, with little channeling, no inelasticity and no isospin-violating interactions, we obtain an improved best fit  ($\chi^2 \sim 100$), corresponding to fitting {\cogent}
compatibly with null experiments,
 %(no channeling is assumed in CDMS-Ge)
but  at the price of giving almost no signal in DAMA,
which prefers a higher DM mass.  We plot this possibility in Fig.\fig{CH}c.

\subsection{Channeling plus inelasticity}
\label{sec:CHinelastic}
The incompatibility between the channeled DAMA and {\cogent} best fits in Fig.\fig{CH}a or c, where we assumed
$f_p/f_n=1$ and $\delta = 0$, can be eliminated varying these two parameters.
Two kind of best fits appear:
\begin{itemize}
\item The `usual' best fit at higher $M_{\rm DM} \approx 6 \GeV$, with a slightly improved $\chi^2$, as shown in Fig.\fig{CH2}a.

\item A new kind of best fit, shown in Fig.\fig{CH2}b,
that makes use of the channeled {\cogent} best fit at smaller $M_{\rm DM}\approx 3\GeV$,
and that employs inelasticity and channeling to also shift the DAMA best fit region to such a small mass.
Indeed, a new local minimum of the DAMA-only fit appears at small mass for large enough channeling in DAMA.  In Fig.\fig{CH2}b we assumed a 20\% channeling fraction, that is enough to have both DAMA best fits with comparable $\chi^2$.
We find the exoDM (shown in the figure) to have a better global fit, however under the assumptions above, a lower mass region exists for endothermic DM  too.
\end{itemize}

The quality of the new best fit is now very significantly improved.
It makes use of  channeling only in DAMA and {\cogent}, of inelasticity and of isospin violation. It relies on channeled sodium events to constitute the bulk of the DAMA signal.

Due to the small fractional modulation (the ratio between the modulated and unmodulated signal) characteristic of exothermic DM, non trivial bounds on this region come from demanding that the DM rate does not exceed the total rate measured by DAMA, in particular in the 1-2 keVee range.
The dark magenta curve in  Fig.\fig{CH2}b shows such bound.
This constraint becomes less stringent for smaller values of $|\delta|$ (which worsen the fit) or under the assumption that the channeling fraction drops at energies below $\sim2$\,keVee.

%
%The reason is that the shift in the DAMA region no longer allows the free parameter $f_p/f_n$ to achieve both its goals described in section~\ref{fpfn}:  getting rid of one experimental bound and obtaining a joint fit of DAMA and {\cogent} at the same time.

%
% different from the one suggested by
%
%dominates for light DM if channeling is negligible.
%If instead channeling is significant, scattering on heavier Iodine dominates, and the region favored by DAMA shifts
%to lower cross sections.
%Fig.\fig{CH}a assumes  a $10\%$ channeling fraction.  The effect is small enough that the best fit is similar to the un-channeled best-fit, with some improvement.  In Fig.\fig{CH}b we consider  a $50\%$ channeling fraction.  Here  a different kind of best-fit is obtained with a different value of $f_p/f_n$,
%but the global fit is not better than in the previous case.

%
%One last remark is in order.  Since both CDMS and {\cogent} use Ge probing a similar range of energies,
%the tension between the two experiments cannot be ameliorated by inventing appropriate DM models.
%A certain amount of channeling in {\cogent} could account for the difference in rates. We  nonetheless ignore this possibility as theoretical calculations shows it to be negligible \cite{channeling} and due to the lack of experimental results on low-temperature Germanium channeling.

\section{Conclusions}\label{concl}
We explored the prospects of dark matter for jointly explaining the modulations signals observed  by DAMA and {\cogent}  while complying with the constraints
from other direct detection experiments: {\sc Xenon10}, \xenon, CDMS-Ge, CDMS-Si and SIMPLE.
In the minimal scenario of spin-independent isospin-symmetric scattering
there is a clear incompatibility, if we adhere to the data analyses performed by the experimental collaborations.  Taking into account uncertainties on the astrophysical velocity distribution of DM and  allowing for a higher sodium quenching factor in DAMA, $q_{\rm Na}=0.4$ does not significantly improve the fits.  Furthermore, DAMA and {\cogent} are mutually incompatible.

Assuming different DM cross sections on proton and neutrons allows to significantly improve the situation, making
DAMA and {\cogent} compatible and weakening one of the bounds from the null experiments.
The best fit is for $f_p/f_n \approx -1.5$, which allows to weaken the dominant bounds from  experiments with Xenon.
However the overall global fit remains  poor, see Fig.\fig{standardf}.
Some additional improvement (although not very significant) is obtained by allowing  for a DM/nucleon form factor that depends on
the transferred momentum and/or on the relative velocity.
A similar results holds for inelastic scattering of DM with a mass splitting, $\delta =M'_{\rm DM}-M_{\rm DM}$.  Both up-scattering ($\delta > 0$) and down-scattering ($\delta<0$) were considered.  The best
 global fit to all relevant experiments is obtained as in Fig.\fig{iDM} for $|\delta| \approx 10$ keV.

A modest channeling in DAMA  allows a further slight improvement.
Allowing also for a $\sim 10\%$ channeling in {\cogent} results in a fit at smaller $M_{\rm DM}\approx 3\GeV$, that, together with
a $\sim20\%$ channeling in DAMA and inelasticity, allows a good global fit (Fig.\fig{CH2}b).

We stress the alternative simple possibility of mild channeling in {\cogent} alone: it allows for a good \cogent\ fit (without assuming a tuned inelasticity nor isospin-violating interactions) but excludes the DAMA signal.

\medskip

In conclusion, we find it hard to explain the full set of current direct detection experiments with spin-independent scattering of dark matter, and contemplate on the possibility that one or both of the positive signals does not arise from DM interactions.

\paragraph{Note added:}  While this work was being finalized, related works appeared~\cite{Belli:2011kw,Schwetz:2011xm}.

\paragraph{Acknowledgements}
We thank Juan Collar, Mariangela Lisanti, Jeremy Mardon and Tracy Slatyer  for useful discussions. We wish to thank the authors of \cite{weinerfit} for pointing out an error in Section 3.2 in an earlier version of the paper.
The work of AS  was supported by the ESF grant MTT8 and by SF0690030s09 project.  The work of DP was supported by the Swiss National Science Foundation under contract No. 200021-116372. The work of MF was supported in part by the European Programme ``Unification in the LHC Era",  contract PITN-GA-2009-237920 (UNILHC).  The work of TV was supported in part by the Director, Office of Science, Office of High Energy and Nuclear Physics, of the US Department of Energy under Contract DE-AC02- 05CH11231.
AS thanks the invitation from the Berkeley physics department, where this work was initiated.
This work was supported by the EU ITN ÒUnification in the LHC EraÓ, contract PITN- GA-2009-237920 (UNILHC).

\end{document}